\documentclass[preprint2]{proto}
\usepackage{natbib}
\bibpunct{(}{)}{;}{a}{,}{,}
\usepackage{times}
\begin{document}

\title{\textbf{\LARGE Early Thermal Evolution of Planetesimals and its Impact on Processing and Dating of Meteoritic Material}}

\author {\textbf{\large Hans-Peter Gail and Mario Trieloff}}
\affil{\small\em Universit\"at Heidelberg}

\author {\textbf{\large Doris Breuer and Tilman Spohn}}
\affil{\small\em Deutsches Zentrum f\"ur Luft- und Raumfahrt, Berlin}

\begin{abstract}
\baselineskip = 11pt
\leftskip = 0.65in 
\rightskip = 0.65in
\parindent=1pc
{\small Radioisotopic ages for meteorites and their components provide constraints on the evolution of small bodies: timescales of accretion, thermal and aqueous metamorphism, differentiation, cooling and impact metamorphism. Realising that the decay heat of short-lived nuclides (e.g. $^{26\!}$Al, $^{60}$Fe), was the main heat source driving differentiation and metamorphism, thermal modeling of small bodies is of utmost importance to set individual meteorite age data into the general context of the thermal evolution of their parent bodies, and to derive general conclusions about the nature of planetary building blocks in the early solar system.
As a general result, modelling easily explains that iron meteorites are older than chondrites, as early formed planetesimals experienced a higher concentration of short-lived nuclides and more severe heating. However, core formation processes may also extend to 10\,Ma after formation of Calcium-Aluminum-rich inclusions (CAIs). A general effect of the porous nature of the starting material is that relatively small bodies ($<$~few~km) will also differentiate if they form within 2\,Ma after CAIs. A particular interesting feature to be explored is the possibility that some chondrites may derive from the outer undifferentiated layers of asteroids that are differentiated in their interiors. This could explain the presence of remnant magnetization in some chondrites due to a planetary magnetic field.
 \\~\\~\\~}

\end{abstract}  

\bigskip
\noindent
\section{\textbf{INTRODUCTION}}
\bigskip

Radioisotopic ages for meteorites and their components provide constraints on the evolution of small bodies: timescales of accretion, thermal and aqueous metamorphism, differentiation, cooling and impact metamorphism. There had been many debates about the nature of the heating source of small bodies, realising that the decay heat of long-lived nuclides (e.g. $^{40}$K, U, Th) is insufficient. Most studies prefer as main heat source driving differentiation and metamorphism the decay heat of $^{26\!}$Al, and possibly $^{60}$Fe, although recent studies \citep{Tan12,Tel12} indicate that its abundance was probably too low to induce major effects.

Thermal modeling of small bodies is of utmost importance to set individual meteorite age data into the general context of the thermal evolution of their parent bodies, and to derive general conclusions about the nature of planetary building blocks in the early solar system.

For undifferentiated chondritic planetesimals, a number of thermal evolution models were constructed that yielded onion-shell type parent body structures with strongly metamorphosed and slowly cooling rocks in deep planetesimal interiors, and weakly metamorphosed fast cooled rocks in shallow outer layers.
Models ranged from mere analytical \citep[e.g.,][]{Miy81,Tri03,Kle08} to numerical models considering initial porosity, sintering or regolith development \citep[e.g.,][]{Ben96,Akr98,Har10,Hen12a,Hen12b,Hen13,Mon13}. There is no final agreement, however, if specific asteroids cooled in an undisturbed onion shell type structure, i.e, some authors argue that this structure was preserved --- at least for a part of the H chondrite asteroid --- as long as 100\,Ma, others argue that this structure was disrupted and reassembled by a large impact event after reaching peak metamorphic temperatures  \citep{Tay87,Sco81,Gan13,Sco13}. However the modeling of onion shell type bodies nevertheless allows insights into thermal evolution without additional complexities involving major collision events.

\begin{table*}[t]

\centerline{\sc Table \refstepcounter{table}\thetable\label{GTBSTabMetTypes}
}
\smallskip
\centering{\sc Properties of important chondrite groups.$^1$
}
\\[2.2ex]

\centerline{\small
\begin{tabular}{@{}lcr@{}lr@{}lr@{}lr@{}lr@{}lrrrrrr@{}}
\hline
\hline
\noalign{\smallskip}
Chondrite group & & EH & & EL & & H & & L & & LL & & CI & CM & CR$^2$ & CO & CK & CV
\\
\noalign{\smallskip}
\hline
\noalign{\smallskip}
Chondrules$^3$                            & [vol\%] & 60-80  & & 60-80  & & 60-80 & & 60-80 & & 60-80 & &  $\ll1$& 20    & 50-60   & 48   & 15      & 45
\\[0.1cm]
Avg. diameter$^3$                         & [mm]    & 0.2    & & 0.6    & & 0.3   & & 0.7   & & 0.9   & & -      & 0.3   & 0.7     & 0.15 & 0.7     & 1.0
\\[0.1cm]
Molten chondr.$^4$                        & [\%]    & 17     & & 17     & & 16    & & 16    & & 16    & & -      & $\approx5$ & $<1$    & 4    & $<1$    & 6
\\[0.1cm]
CAIs$^3$                                  & [vol\%] &   0.1-1&?&   0.1-1&?&  0.1-1&?&  0.1-1&?&  0.1-1&?& $\ll1$ & 5     & 0.5     & 13   & 4       & 10
\\[0.1cm] 
Matrix$^3$                                & [vol\%] & $<2$-15&?& $<2$-15&?& 10-15 & & 10-15 & & 10-15 & & $>99$  & 70    & 30-50   & 34   & 75       & 40
\\[0.1cm]
FeNi metal$^{3,5}$                        & [vol\%] &       8&?&      15&?& 10    & & 5     & & 2     & & 0      & 0.1   & 5-8     & 1-5   & $<0.01$ & 0-5
\\[0.1cm]
Fe$_\mathrm{metal}$/Fe$_\mathrm{total}^6$ &         & 0.76   & & 0.83   & & 0.58  & & 0.29  & & 0.11  & & 0      & 0     & & 0-0.2 &       & 0-0.3
\\[0.1cm] 
Fayalite in Olv$^6$                       & [mol\%] &        & &        & & 16-20 & & 23-26 & & 27-32 & &
\\[0.1cm]
Mg/Si$^7$                                 &         & 0.75   & & 0.85   & & 0.95  & & 0.93  & & 0.94  & & 1.05   & 1.05  & 1.06    &  1.06 & 1.08    &  1.05
\\[0.1cm]
Al/Si$^7$                                 &         & 0.051  & & 0.055  & & 0.065 & & 0.065 & & 0.065 & & 0.085  & 0.093 & 0.079   & 0.092 & 0.097   & 0.111
\\[0.1cm]
Ca/Si$^7$                                 &         & 0.036  & & 0.038  & & 0.050 & & 0.050 & & 0.049 & & 0.061  & 0.071 & 0.060   & 0.700 & 0.075   & 0.082
\\[0.1cm]
Fe/Si$^7$                                 &         & 0.92   & & 0.66   & & 0.80  & & 0.59  & & 0.53  & & 0.86   & 0.84  & 0.80    &  0.80  & 0.73   & 0.75
\\[0.1cm]
(Refract./Mg)$^8_\mathrm{CI}$             &         & 0.87   & & 0.83   & & 0.93  & & 0.94  & & 0.90  & & 1.00   & 1.15  & 1.03    & 1.13  & 1.21    & 1.35
\\[0.1cm]
Petrol. types$^6$                         &         & 3-5    & & 3-6    & & 3-6   & & 3-6   & & 3-6   & & 1      & 2     & 2       & 3     & 3-6     & 3
\\[0.1cm] 
Max. temp.$^6$                            & [K]     & 1020   & & 1220   & & 1220  & & 1220  & & 1220  & & 430    & 670   & 670     & 870   & 1220    & 870
\\
\noalign{\smallskip}
\hline
\end{tabular}
}

\bigskip\noindent
\parbox{0.99\hsize}{\scriptsize 
Notes: 
(1) Adapted from \citet{Tri06}, with modifications. (2) CR group without CH chondrites. (3) From \citet{Pap98}.  (4) From \citet{Rub95}. (5) In matrix. (6) From \citet{Sea88}. (7) From \citet{Hop08}. (8) From \citet{Sco96}.
}

\end{table*}

For differentiated planetesimals numerical models were developed to describe, e.g., the differentiation of Vesta \citep{Gho98}. Important processes that need to be considered are formation time, but also duration of accretion \citep{Mer02} convection and melt migration \citep{Hev06,Sah07,Mos11,Neu12}. Only few models take into account initial porosity, sintering, the accretion process and redistribution of heat sources during core and mantle/crust formation. Chronological constraints are derived from iron meteorites \citep[e.g.,][]{Kle05,Kle09} and basaltic achondrites \citep[e.g.,][]{Biz05}.

\bigskip
\noindent
\textbf{\refstepcounter{subsection}\thesubsection\ Meteorites and their components}

\bigskip

Much of our knowledge about the formation of first solids and planetary bodies in the solar system is derived from studies of extraterrestrial matter. Meteorites were available for laboratory investigations long before extraterrestrial samples returned by space probes. Most meteorites are from asteroids, which in turn are the remnants of a swarm of planetesimals leftover from the planet formation process 4.6 Ga ago. The diversity of meteorite classes available in worldwide collections implies that they are derived from about 100 different parent bodies. Only a minority of meteorites are from asteroids that were differentiated like the terrestrial planets, with metallic Fe-Ni cores, silicate mantles and crusts. These meteorites comprise various classes of iron meteorites, and meteorites of mainly silicate composition (e.g., basaltic, lherzolitic, etc.). The largest group of differentiated silicate meteorites belong to the HED group (Howardites, eucrites, diogenites) and are commonly assumed to stem from Vesta, based on dynamical reasons and spectroscopy (see section 6). Another important group of basaltic meteorites are the angrites, for which no parent body has been assigned conclusively. Iron meteorites and basaltic achondrites yield surprisingly young radiometric ages \citep{Tri09}, consistent with the picture that they formed and differentiated early in solar system history. This evidences that there was a heat source providing differentiation energ even for small bodies in the early solar system.

On the other hand, the majority of meteorites is undifferentiated. Such meteorites have to a first order solar abundances of Fe, where Fe is partly present as metal, and partly oxidised residing in silicates (see Table~\ref{GTBSTabMetTypes}). The term ``undifferentiated'' also refers to the simultaneous existence of phases formed at low temperatures --- represented by fine-grained matrix rich in volatile elements --- and phases formed at high temperature: Prominent examples are rare, cm-sized Calcium-Aluminum-rich inclusions (CAIs) and abundant sub-mm to mm-sized droplets, crystallized from liquid silicates, that are called chondrules --- after which undifferentiated meteorites are named chondrites. These tiny objects formed by yet unknown flash heating processes in the solar nebula, were rapidly cooled (order of K/hour or K/day) and predate accumulation of planetesimals .

\begin{table*}[t]

\centerline{\sc Table \refstepcounter{table}\thetable\label{GTBSTabExtinctNuc}
}
\smallskip
\centering{\sc Radioactives important for thermochronology and heating of planetary bodies.
}
\\[2.2ex]
{\small
\begin{tabular}{r@{}l r@{}l r@{$\;\times\;$}l r@{$\;\times\;$}l r@{}l r@{$\;\times\;$}l r@{$\;\to\;$}ll}
\hline
\hline
\noalign{\smallskip}
\multicolumn{2}{c}{Parent}  & \multicolumn{2}{c}{Daughter} &  \multicolumn{2}{c}{Half-life$^{\rm a}$} & \multicolumn{2}{c}{Abundance$^{\rm b}$} & \multicolumn{2}{c}{Reference}  & \multicolumn{2}{c}{Heating rate} & \multicolumn{2}{c}{Main decay modes} & percent
\\[.2cm]
  \multicolumn{2}{c}{}  &   \multicolumn{2}{c}{} &  \multicolumn{2}{c}{a} &  \multicolumn{2}{c}{}  &  \multicolumn{2}{c}{}  &  \multicolumn{2}{c}{W/Kg}        &  \multicolumn{2}{c}{}  \\
\noalign{\smallskip}
\hline
\\[-0.7ex]
$^{26\!}$&Al & $^{26}$&Mg & $7.17$&$10^5$ & $5.2$&$10^{-5}$ & $^{27\!}$&Al & $1.67$&$10^{-7}$ & $^{26}$Al & $^{26}$Mg$+\beta^++\nu$ & 82 
\\[-.0cm]
   \multicolumn{2}{c}{}  &   \multicolumn{2}{c}{}  &  \multicolumn{2}{c}{}  &  \multicolumn{2}{c}{}  &  \multicolumn{2}{c}{}  & \multicolumn{2}{c}{}  & $^{26}$Al + e & $^{26}$Mg$$ + $\bar\nu$ & 18 
\\[.1cm]
$^{60}$&Fe & $^{60}$&Ni  & $2.6$&$10^6$  & $5.8$&$10^{-9}$ & $^{56}$&Fe  & $2.74$&$10^{-8}$ & $^{60}$Fe & $^{60}$Co + $\beta^-$ + $\nu$ & 100 
\\[-.0cm]
  \multicolumn{2}{c}{}  &   \multicolumn{2}{c}{} & $5.27$&$10^0$  &   \multicolumn{2}{c}{} &   \multicolumn{2}{c}{}  &  \multicolumn{2}{c}{}  & $^{60}$Co & $^{60}$Ni + $\beta^-$ + $\nu$ & 100
\\[.1cm]
$^{53}$&Mn & $^{53}$&Cr  & $3.7$&$10^6$ & $6.3$&$10^{-6}$  & $^{55}$&Mn  &  \multicolumn{2}{c}{}  & $^{53}$Mn + e & $^{53}$Cr + $\bar\nu$ & 100
\\[.1cm]
$^{182}$&Hf & $^{182}$&W & $8.9$&$10^6$ & $9.7$&$10^{-5}$ & $^{180}$&Hf &  \multicolumn{2}{c}{}  &  $^{182}$Hf & $^{182}$Ta + $\beta^-$ + $\bar\nu$ & 100  \\
  \multicolumn{2}{c}{} &  \multicolumn{2}{c}{} &   \multicolumn{2}{r}{114.43 d\ \,\,}  &   \multicolumn{2}{c}{}   &   \multicolumn{2}{c}{}   &  \multicolumn{2}{c}{} &   $^{182}$Ta & $^{182}$W + $\beta^-$ + $\bar\nu$ & 100
\\[.1cm]
$^{129}$&I & $^{129}$&Xe &  $1.6$&$10^7$ &  $1.2$&$10^{-4}$ & $^{127}$&I &  \multicolumn{2}{c}{} & $^{129}$I & $^{129}$Xe + $\beta^-$ + $\bar\nu$ & 100  
\\[.1cm]
$^{244}$&Pu & $^{131-136}$&Xe &  $8.2$&$10^7$ &  $6.6$&$10^{-3}$ & $^{238}$&U & \multicolumn{2}{c}{} & $^{244}$Pu &  spont. fission
\\[.1cm]
$^{40}$&K  & $^{40}$&Ca &  $1.26$&$10^9$ &  $1.58$&$10^{-3}$  &  $^{39}$&K &  $2.26$&$10^{-11}$ & $^{40}$K  & 
$^{40}$Ca + $\beta^-$ + $\bar\nu$ & 89 \\
   \multicolumn{2}{c}{}  & $^{40\!}$&Ar &  \multicolumn{2}{c}{}  &  \multicolumn{2}{c}{} &  \multicolumn{2}{c}{} & \multicolumn{2}{c}{} &  $^{40}$K + e & $^{40}$Ar + $\bar\nu$ & 10.7
\\[.1cm]
$^{232}$&Th &  $^{208}$&Pb    & $1.401$&$10^{10}$ &  $4.40$&$10^{-8}$ &  &Si   & $1.30$&$10^{-12}$ & $^{232}$Th & $^{208}$Pb + 6$\alpha$ + 4$\beta$  &
\\[.1cm]
$^{235}$&U  &  $^{207}$&Pb    & $7.038$&$10^8$ &  $5.80$&$10^{-9}$  & &Si      & $3.66$&$10^{-12}$ & $^{235}$U & $^{207}$Pb + 7$\alpha$ + 4$\beta$ &
\\[.1cm]
$^{238}$&U  &  $^{206}$&Pb    &  $4.468$&$10^9$ & $1.80$&$10^{-8}$  & &Si      & $1.92$&$10^{-12}$ & $^{238}$U & $^{206}$Pb + 8$\alpha$ + $7\beta$ &  \\
\noalign{\smallskip}
\hline
\end{tabular}
}

\renewcommand{\baselinestretch}{1.0}\scriptsize
\bigskip\noindent
\parbox{1.\hsize}{\scriptsize 
Notes:  (a) Half-lives are from \citet{Rug09} for $^{60}$Fe, from \citet{Mat09} for $^{26}$Al, (b) Abundances for alive nuclei at formation time of the solar system are from \citet{Lod09}, for extinct nuclei for $^{26\!}$Al from \citet{Jac08}, $^{60}$Fe from \citet{Tan12}, $^{53}$Mn from \citet{Tri08}, $^{182}$Hf from \citet{Bur08}, $^{129}$I from \citet{Bra99}, $^{244}$Pu from \citet{Hud89}.
}

\end{table*}

The two major classes of undifferentiated meteorites are ordinary chondrites --- the most abundant class of meteorites delivered to Earth --- and carbonaceous chondrites. Carbonaceous chondrites can have significant amounts of water and carbon and contain more matrix and fewer chondrules. Matrix often contains hydrous
minerals resulting from ancient interaction of liquid water and primary minerals, indicating mild thermal heating of the parent body initiating fluid flow.

In general, chondrites represent undifferentiated, chemically ``primitive'' material with approximately solar element abundances of non-volatile elements. However, the various chondrite groups have characteristic compositional differences, e.g., in major element ratios like Mg/Si and Al/Si, or oxygen isotopic composition.
A further well known classification criterion is the strong variation in the abundance of Fe in reduced (metallic) and oxidised form, i.e., the Fe content of
silicates.

The most reduced groups are EH and EL enstatite chondrites, named after the Fe-free Mg-endmember enstatite of the mineral group pyroxene. Here, Fe is not present in oxidised form in silicates, it occurs mainly in metals and sulfides. Ordinary chondrites are classified according to the total abundance of Fe, and the ratio of reduced (metallic) Fe to oxidised Fe: H (High total iron), L (Low total iron) and LL (Low total iron, Low metal). Carbonaceous chondrites are classified according to their contents of volatile elements, but also other parameters. They are named after a prominent member of the respective group: CI (Ivuna), CM (Mighei), CO (Ornans), CK (Karoonda), CV (Vigarano), CR (Renazzo). Some of the characteristic properties of these meteorite groups are collected in Table~\ref{GTBSTabMetTypes}.

Based on several lines of reasoning, each chondrite group and its unique composition is assigned to a specific independent parent body. Although they remained undifferentiated, chondritic rocks were heated to variable peak metamorphic temperatures described by the so called petrologic type. In general, petrologic type 3 represents the most pristine, unaltered material. In the case of (essentially dry) ordinary chondrites, petrologic types 4 to 6 describe a sequence of increasing metamorphism, at temperatures up to 1\,200 --- 1\,300\,K. In the case of water bearing carbonaceous chondrites, the sequence of type 3 - type 2 - type 1 is a sequence of increasing intensity of aqueous metamorphism and increasing alteration by low temperature fluids \citep[see][for details of classification]{Weis06}.

The existence of different petrologic types among individual chondrite groups demonstrates that different heating effects and/or cooling conditions prevailed on
individual parent bodies. Thermal modeling should explain these variable heating and cooling scenarios reflected by chondritic rocks.

\bigskip
\noindent
\textbf{\refstepcounter{subsection}\thesubsection\ Radioactives in the early solar system}
\bigskip

A number of now-extinct short-lived and long-lived radioactive nuclei were present in the early solar system, that play a key role for heating of planetesimals and for dating key processes during formation of the solar system \citep[see, e.g.][ for a review]{Kee03,Kro09,Dau11}. Table \ref{GTBSTabExtinctNuc}  gives an overview of some important radioisotopes present in the early solar system, their daughter nuclides, decay modes and their initial abundances at time of CAI formation. Radioactive isotopes are important for early solar system evolution in two respects:

1. Radioisotopes can be used to trace the cooling histories of meteorites and their parent bodies via isotopic dating methods (see section \ref{GTBSSectThermChron}). While long-lived nuclides like $^{238}$U, $^{235}$U, and $^{40}$K are widespread used in geochronology (U-Pb-Pb and Ar-Ar dating), short-lived nuclides with half life's of $<100$\,Ma like $^{26\!}$Al, $^{53}$Mn, $^{182}$Hf, $^{129}$I or $^{244}$Pu
are mainly used for meteorites which formed within the first tens of millions years of the solar system. Ages based on one specific short-lived nuclide define only relative time scales that need to be calibrated to an absolute time scale (typically by fast cooled precisely dated rocks called tie points). However, the chronologic interpretation of initial isotope abundances as ``ages" requires a homogeneous distribution of the parent nuclide in the solar nebula, which is for some cases still a matter of debate.

2. Radioisotopes were furthermore an important heat source for early formed planetesimals and caused aqueous and thermal metamorphism, or even melting and differentiation. Particularly $^{26\!}$Al was present with an abundance which should have severe heating effects on early formed bodies of bigger than a few km in size (see  section~\ref{GTBSAnalytSol}).

\begin{figure}[t]
\epsscale{0.85}

\centerline{
\plotone{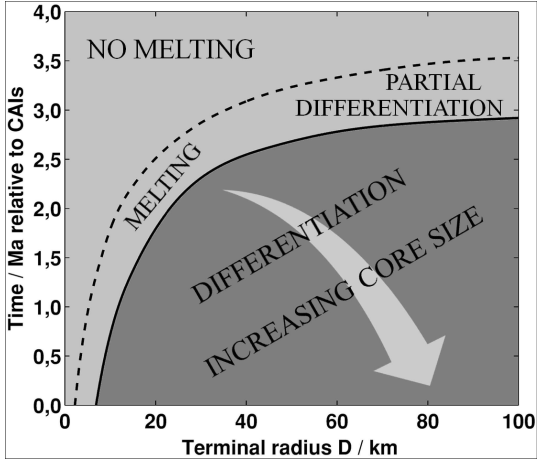}
}

\caption{\small Formation time (assuming instantaneous accretion) as a function of the terminal radius of a planetesimal indicating the  parameter space for which a planetesimal forms a core (dark grey), melting occurs but no core differentiation (light grey area in between solid and dashed line) and no melting occurs (light grey area above dashed line) \citep[from][]{Neu12}.   }

\label{GTBSFigRTaccDiffrent}
\end{figure}

\bigskip
\centerline{\textbf{\refstepcounter{section}\thesection. INTERNAL CONSTITUTION AND THERMAL\label{GTBSSectInternConst}
}}
\centerline{\textbf{EVOLUTION OF PLANETESIMALS
}}
\bigskip

The evolution of small planetary bodies such as asteroids begins with its accretion from the protoplanetary dust and ice as a porous aggregate \citep[e.g.][ see also chapters by Johansen et al., and Raymond et al.]{Mor12}. Heating by radioactive decay combined with the self-gravity leads to the substantial compaction at sub-solidus temperatures and in case of icy bodies to crystallization of amorphous ice and phase transitions of volatile species. Provided strong radiogenic heating by $^{26\!}$Al and $^{60}$Fe, the iron and silicate phases start to melt upon reaching the respective solidus temperature --- heating by impacts has played a minor role for melting on bodies smaller than 1000\,km \citep{Kei97,Sra12}. If melting is wide spread enough, the planetary body can differentiate (at least partially) into an iron core and a silicate mantle. Differentiation is initiated by the iron melt sinking downward and silicate melt percolating towards the surface. In case of icy bodies, melting of the ice starts before iron and silicates and thus we have a third component that is possibly separated. Whether a small planetary body remains undifferentiated or starts to melt and differentiate depends mainly on the composition, the accretion time that is associated with the available amount of short-lived radioactive elements and the size of the body. The earlier the onset of accretion and the larger the body, the higher are the interior temperatures and the more likely is the differentiation (see Fig~\ref{GTBSFigRTaccDiffrent}). In the following we discuss in more detail the evolution and interior structure of the different types of small planetary bodies, i.e., undifferentiated, differentiated and icy bodies.

\begin{figure}[t]
\epsscale{1.0}

\centerline{
\plotone{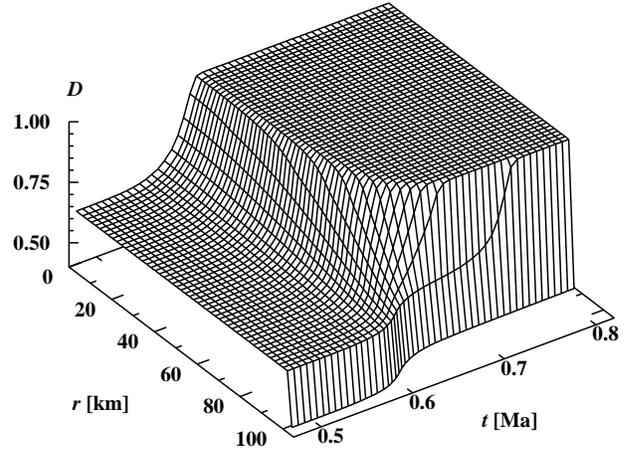}
}

\caption{\small Evolution of the filling factor $D$ by sintering of a body with  initially 100\,km radius and an inital filling factor of $D=0.64$ (= random densest packing of equal sized spheres). At about one half-life ($\tau_{1/2}$) of $^{26\!}$Al  after (instantaneous) formation of the body there is a sudden shrinking of the body due to onset of sintering \citep[from][]{Hen12a}.}
\label{GTBSFigSinterPor}
\end{figure}

\bigskip
\noindent
\textbf{\refstepcounter{subsection}\thesubsection\ Undifferentiated bodies
}
\bigskip

In the initial state, the planetesimals are assumed to be highly porous as they initially grew as aggregates of dusty and unconsolidated material of the protoplanetary discs. As most meteorites are well-consolidated \citep{Bri03}, the planetesimals have experienced some form of compaction from the unconsolidated and highly porous to consolidated state. Two subsequent stages of compaction are generally assumed: cold compaction simply due to self-gravitation \citep[e.g., ][]{Gue09} and hot pressing, i.e., the so-called sintering, which is associated with the deformation of the crystals due to temperature and pressure \citep{Rao72,Yom84}. The surface porosity is further influenced by impact processes that also result in a compaction of the dusty material \citep{Wei09}. For planetary bodies bigger than 10\,km most part of the interior is already compacted by cold isostatic pressing to about a density close to that of the densest random packing of equal-sized spheres. The porosity $\phi$ (i.e., the fraction of voids per unit volume) of the closest random packing is about 0.36--0.4 \citep{Son08, Hen12a}. In the subsequent evolution when the planetary body is heated by the radioactive decay, sintering starts in the deep interior where pressure and temperature are highest. The threshold temperature, at which sintering becomes active and the porosity rapidly approaches zero, is about 700 -- 750\,K assuming the simple treatment of sintering from \citet{Yom84}. With further heating, the compacted region moves toward the surface and the planetary body shrinks in size as the porosity (volume) is decreasing. Figure~\ref{GTBSFigSinterPor} shows a typical evolution of the radial distribution of the filling factor $D$ (defined as $D=1-\phi$, note that in part of the literature the filling factor $D$ is denoted as $\phi$). Assuming an initial porosity of the planetary body between 40 and 60\% the radius shrinks by about 15 -- 25\% of its initial value in about 0.6 decay timescales of $^{26\!}$Al. 

A porous layer remains at the surface as the threshold temperature of sintering cannot be reached there. The thickness of this porous layer varies with the onset time of formation and the size of the planetary body --- the later the onset time and the smaller the body, the less heat is provided by radioactive decay and the larger is the relative thickness of the outer porous regolith layer. An important effect of sintering is the associated change in the thermal conductivity of the material. The thermal conductivity is about two to three orders of magnitude higher for compacted material in comparison to highly porous material \citep[e.g., ][see also section \ref{GTBSProMod}]{Kra11b,Kra11}. Various thermal evolution models for planetesimals \citep{Yom84,Sah07,Gup10,Hen12a,Neu13} have demonstrated the importance of sintering and the associated change of the thermal conductivity: the porous layer thermally insulates the interior resulting in temperatures in the outer layers that increase more rapidly than in the models assuming a homogeneous body without porosity changes \citep{Miy81}. This observation has in particular consequences on the burial depths of meteorites and predicts shallower outer layers for the petrologic types (see section~\ref{GTBSSectReconstr}).  It should be noted that sintering as well as the rate of change of the interior temperatures are dampened by any kind of non-instantaneous accretion \citep{Mer02,Neu12}. Hence the thickness of the porous regolith layer increases with the duration of accretion.

The typical cooling history shows that the planetary body is first heated by decay of $^{26\!}$Al and possibly $^{60}$Fe and then cools down over an extended period.  The temperature distribution in an undifferentiated but sintered planetesimal shows a typical onion-like thermal structure for which the temperature decreases from the center toward the surface \citep{Ben96,Har10,Hen12a,Hen12b,Mon13}. The coupling of accretion and sintering, however, leads to an episodic evolution of the central peak temperature with at least two heating phases during the first few Ma \citep{Neu12}.  The heating by the long-lived isotopes of U, Th and K becomes significant only after $\approx8$\,Ma after CAIs. These nuclides do not provide an effective heat source because in asteroid-sized bodies their heat generation rate is of the same magnitude as the rate of heat loss through the surface \citep{Yom84}. The temperature distribution is consistent with an onion-shell structure in which the degree of thermal metamorphism is a function of depth within the parent body. The time scale of the heating and cooling phase depends on the size of the body and its formation time --- thus results from thermal evolution models can be used to fit the chronological data of meteorites to obtain the onset time of formation, the size of the parent body and the burial depths of the meteorites (see sections \ref{GTBSSectThermChron} and~\ref{GTBSSectReconstr}).

\begin{figure*}[t]
\epsscale{1.0}

\centerline{
\includegraphics[width=\hsize]{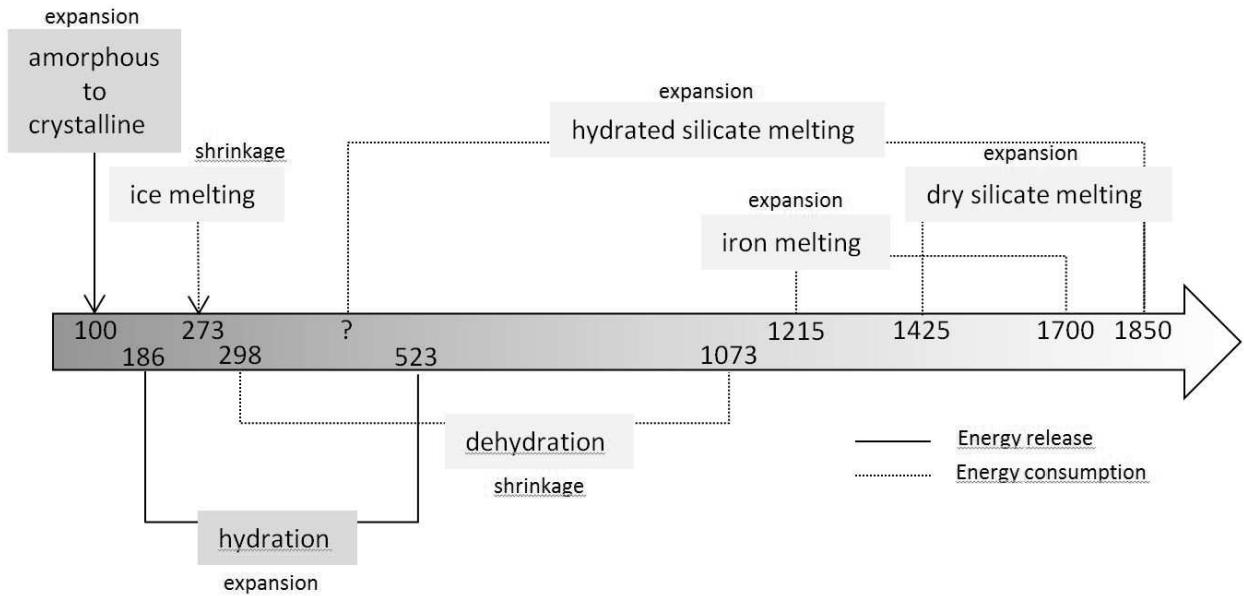}
}

\caption{\small Sketch showing processes that occur at certain temperatures or temperature ranges in icy planetesimals. Numbers are temperature in Kelvin, note that absolute values will vary with the assumed composition. Considering that a body heats up with time during and after its accretion, these processes change subsequently during the thermal evolution and influence the temperature and the structural evolution of a planetesimal. Note that when the body cools after the initial heating phase, energy will be released by crystallization of melt and the volume changes are reversed. 
}
\label{GTBSFigtimeline}
\end{figure*}

\bigskip
\noindent
\textbf{\refstepcounter{subsection}\thesubsection\ Differentiated bodies\label{GTBSSectDiffer}
}
\bigskip

In contrast to undifferentiated bodies, melting has occurred in the interior of differentiated bodies and processes like heat transport via magma segregation or convection and redistribution of the radioactive isotopes play an important role in their thermo-chemical evolution --- effects that also tend to dampen the interior temperature increase. As in the case of the undifferentiated bodies, in the initial stage the differentiated bodies accrete as porous material and compacts due to cold compaction and hot pressing. Melting starts when the solidus temperature of the iron and silicate phase is reached. Typically the iron-rich phase starts to melt at a lower temperature than the silicates for a volatile poor body. 

It has been shown that the heat production by the short-lived radioactive nuclei coupled with an initially porous state suffices to achieve melting even in a small planetesimal with the radius of only a few km in case they form simultaneously with the CAIs \citep{Hev06,Mos11,Hen12a,Neu12} (Fig.~\ref{GTBSFigRTaccDiffrent}). With increasing formation time a larger planetesimal is required for melting to occur. After a formation time of about 2.8 -- 3\,Ma melting and differentiation is unlikely assuming a planetesimal with a composition consistent to ordinary chondrites \citep[e.g., ][]{Gho98,Sah07,Mos11,Neu12}. This threshold may shift to later formation times if the material consists of volatiles reducing the melting temperature. 

The occurrence of melt does not necessarily result into the differentiation (core-mantle or mantle-crust differentiation) of the planetary body. Whether partial melt can separate from the solid matrix depends on the density contrast between the molten and the solid material, the amount of partial melt and the melt network. In particular, an interconnected melt network is required as melt will not separate from the residual solid matrix if trapped in disconnected pockets. These pockets occur when the so called dihedral angle exceeds value of 60$^\circ$ and the melt fraction stays below a certain critical melt fraction \citep{VBa86}. For a dihedral angle smaller than 60$^\circ$, a stable interconnected melt channel occurs for all melt fractions. In most partial melts including basic Si melts, the dihedral angle is less than 60$^\circ$ \citep{Bee75,Waf79}. Thus, even if the silicates melt fraction is as small as a few percent, an interconnected network forms and is able to segregate \citep{Tay93,Maa03}. For iron melt the situation is more controversial: Experimental studies --- most at higher pressures than expected for melting in small planetary bodies --- have shown that partial melting does not lead to metal melt migration unless a substantial fraction of partial melt is present \citep[e.g., ][]{Tak83,Wal88,Gae99}. Only recent experiments suggest an interconnected melt network for pressures below 2 -- 3 GPa \citep{Ter08}. Furthermore, Fe,Ni-FeS veins have been identified in the acapulcoites and lodranites, suggesting melt migration for low degree of melting \citep{MCo96}. However, these veins represent only short distance migration suggesting that sulfide-rich early partial melt was not separated from the metal-rich residue \citep{MCo97}. 

Different scenarios and relative timing between the core-mantle and mantle-crust differentiation have been proposed of which the following three main scenarios can be found in the literature: Iron segregation and core formation occurs already for a small melt fraction and even before silicate starts to melt \citep{Hew88,Sah07}. Thus, mantle-crust differentiation follows core formation. In contrast to the first scenario, it is suggested that core formation requires a substantially larger melt fraction including melt of the silicates. Assuming a melt threshold of about 40\%, mantle-crust differentiation starts before core formation with an onset time of melt extrusions between 0.15 -- 6\,Ma after the CAIs \citep{Gup10}. In the third scenario, a threshold of more than 50\% melt is assumed for metal to efficiently segregate into the center \citep[e.g., ][]{Tay92,Tay93,Rig97,Dra01,Hev06,Gup10,Mos11} --- thus, the presence of an early magma ocean is required to form a core. These models further imply that core formation is almost instantaneous and is followed --- as in the first scenario --- by the mantle-crust differentiation. The existence of an early magma ocean on planetesimals that form with or shortly after the CAIs has been also suggested to explain the observed remnant magnetization in meteorites \citep{Wei08b,Car10,Wei10,Fu12}. It should be noted though that in most models studying the influence of differentiation on the thermal evolution of a planetesimal a certain scenario is prescribed, i.e, segregation of either silicates or iron occurs at a fixed melt fraction and is not calculated self-consistently. 

A first attempt to model the differentiation process self-consistently via porous flow by \citet{Neu12} indicates that the timing of core formation with respect to the occurrence of silicate melting also depends on the amount of  light elements in the iron phase, which lower the melting temperature, such as sulfur: For a sulfur rich composition core formation is prior to silicate melting similar to the findings by \citet{Hew88} but for a low sulfur content as in fact suggested for H-chondrites \citep{Kei62} core formation starts simultaneously with silicate melting but is completed earlier than the mantle crust differentiation. This difference in timing is caused by the variation in the migration velocity through porous media that depends among others on the melt fraction. In the case of low sulfur content the temperature difference between solidus and liqui\-dus of the Fe-FeS mixture is a few hundred degrees whereas it is only a few tens of degrees for a high sulfur content. Thus, assuming the same temperature increase by radioactive heat sources and the same mass of the entire iron rich phase, a smaller melt fraction is available for low sulfur content, which then migrates on timescales longer than the mean life of the short lived radioactive isotopes --- the temperature and the amount of melting increase before melt segregates. As a further consequence, core formation takes more than 2\,Ma and up to 10\,Ma for planetesimals that experience less than 50\% degree of melting.

As mentioned above, for degrees of melting higher than about 50\%, a magma ocean is present in a planetesimal. At this melt threshold a change in the matrix structure can be observed that is associated with a strong decrease of the viscosity to values representing the liquid material ($\approx0.1$ -- 1 Pas). In this region of high degree of melting, the melt and heat transport is not via porous flow but vigorous convection can set in to cool efficiently the interior. A first numerical study to model the convective cooling of a magma ocean is by \citet{Hev06}. They incorporate increased heat transfer by convection once the melt fraction exceeds a value of about 50\%. Due to efficient cooling, further heating by $^{26\!}$Al and $^{60}$Fe does not increase the magma ocean temperature and thus the degree of partial melting, but increases the extent of the magma ocean  --- the planetesimal becomes a partly molten sphere undergoing vigorous convection below a thin rigid shell. A similar process has been suggested by \citet{Wei08b} and \citet{Elk11} who show that the efficient cooling of the magma ocean might be responsible for the generation of a magnetic field in the iron-rich core of parent bodies lasting more than 10\,Ma. These studies, however, neglected the partitioning of $^{26\!}$Al into the silicate melt. Considering that migration of $^{26\!}$Al enriched melt towards the surface dampens strongly the internal temperatures even before a large scale internal magma ocean can be generated \citep{Hev06,Mos11,Neu13}.  Precluding a global magma ocean is supported by  \citet{Wil12} who suggest that volatile rich magmas in the mantles would be removed very efficiently and only small amounts of melt in the interior would be present at any time. 

Although details in the timing of the differentiation processes differ between the various models, most thermo-chemical models suggest that  both chondritic meteorites and differentiated meteorites may originate from one and the same parent body \citep{Sah05,Biz05,Hev06,Sah07,Wei10,Elk11,Neu12,For13}. These models show that a typical structure consists of a differentiated interior with an iron core and a silicate mantle that is covered by an undifferentiated upper layer with its lower part being sintered and compacted and its upper part still porous.  An interior structure that may also explain the remnant magnetisation of the CV meteorite Allende \citep{Wei10} --- although the origin of the magnetization is controversially discussed \citep{Bla11} --- and can develop as the basaltic magmas rising by porous flow cool and solidify while rising toward the surface, leaving the upper layer unaffected  \citep{Elk11,Neu12}. Furthermore, Lutetia with its high bulk density of 3400 $\pm$ 300 kgm$^{-3}$ \citep{Pae11}  and its primitive surface could also be a potential candidate \citep{Wei12,For13a,Neu13a}. An undisturbed primordial surface layer is, however, more difficult to preserve for high degree of partial melting (e.g., early formation relative to CAIs) and further considering the enrichment of  $^{26\!}$Al into the melt \citep{Neu13}, in case of an increase in melt buoyancy due to exsolution of volatiles \citep{Wil91,Kei93} and/or a decrease in the resistance to flow due to coalescence of melt channels into larger veins and dikes \citep{Wil08} --- processes that also tend to preclude a deep magma ocean (see above). Instead, a much more efficient transport of melt toward the surface and a `destruction' of the upper primordial layer is likely. In the work by \citet{Wil91}, the authors further argue that basaltic melt that contains more than a few hundred parts per million of volatiles is lost to space by explosive volcanism for parent bodies of  less than 100\,km in radius. This process might explain the lack of observed asteroids with a basaltic crust. For larger bodies such as Vesta, which is one of the few asteroids showing a basaltic crust at its surface \citep{Bin97,DeS12}, explosive silicate eruptions can not escape the gravity field and remain at the surface. In conclusion, an early formation time relative to the CAIs and/or a volatile rich composition favours the existence of a basaltic crust (unless it is lost to space), whereas a late formation time and a dry composition is in more in favour of an undifferentiated layer above a differentiated interior.

\begin{figure*}[t]
\epsscale{1.0}

\centerline{
\includegraphics[width=\hsize]{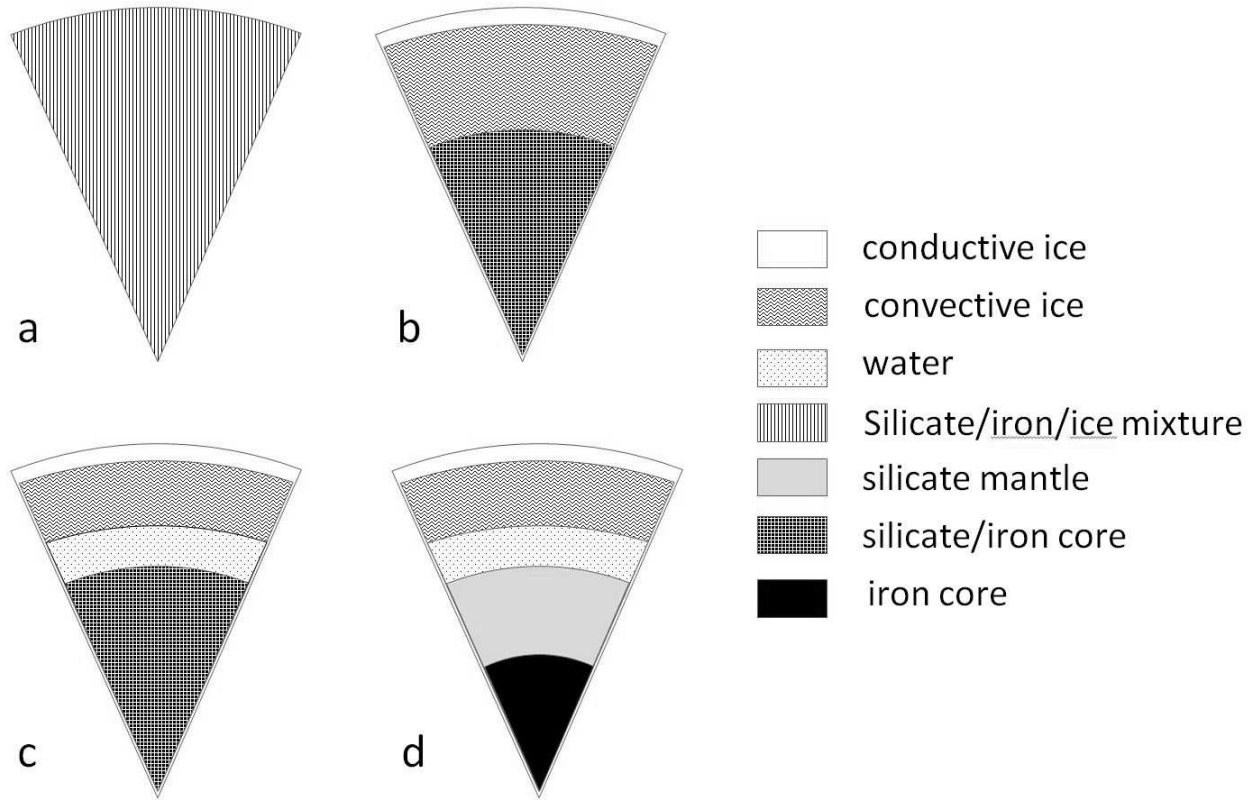}
}

\caption{\small Structural end members of small icy bodies based on different evolution scenarios: (a) homogeneous body made of a mixture of H$_2$O and high-density silicates, (b) differentiated body with high-density silicate core or low-density serpentine and outer ice layer, (c) same as (b) but the presence of antifreeze material (ammonia) maintains a liquid layer, and (d) fully differentiated model with an inner iron core \citep[adapted from][]{McC05}.
}
\label{GTBSFigstructure}
\end{figure*}

\bigskip
\noindent
\textbf{\refstepcounter{subsection}\thesubsection\ Icy bodies
}
\bigskip

The thermal and structural evolution of a planetesimal can vary substantially depending on the concentration of water. The presence of water ice may moderate the temperature increase during heating as substantial amounts of heat may be absorbed during melting. Moreover, the resulting warm melt water may  redistribute heat within the body by hydrothermal convection and by porous flow  \citep{Gri89}. In addition, other processes that are not common of dry silicate bodies may influence the thermal evolution of icy bodies. Among these are exothermic processes such as the crystallization of amorphous ice (one of the major possible morphological transformations inside small icy planetary bodies, according to \citet{Pri92}, \citet{Pri99}, the formation of hydrous silicates (e.g. serpentine), as well as endothermic dehydration reactions. The release of latent heat through hydration reactions, in particular, may cause runaway melting of ice as has been demonstrated by \citet{Coh00}. The initiation of the above processes depend mostly on temperature and occur at times that are determined by the thermal evolution of the planetesimal. They are usually associated with density changes in the interiors. Expansion or shrinkage of the planetesimals due to the density changes may even shape the surfaces of these bodies (see~Fig.~\ref{GTBSFigtimeline}). The following list details the processes in order of occurrence during heating of an icy  planetesimal:

1. The exothermic transition from amorphous to crystalline ice occurs at about 100\,K \citep{Mer03,Mer06}, thereby the density of ice (940\,kg\,m$^{-3}$ for amorphous ice) decreases, causing expansion and the thermal conductivity of the ice increases by a factor of $\approx20$. 

2. Water ice melting is an endothermic reaction and during the melting of ice the density increases to 1\,000\,kg\,m$^{-3}$ causing shrinkage of the body. However, with further increasing temperature up to the evaporation temperature, the density decreases again and results in an expansion. Ice melting is also related to hydrothermal convection and differentiation of ice, water, and silicates and the associated heat transport.

3. Hydration of silicate minerals (e.g., the formation of serpentine from forsterite and H$_2$O) can occur simultaneously with ice melting but across a larger temperature interval. Hydration reduces the density of the rock and is an exothermic reaction. 

4. The dehydration of silicates occurring at higher temperatures is endothermic and consumes more energy than is produced by hydration. Release of the water bonded in minerals increases the density of the rock phase and causes shrinking of the body.

5. Silicate and iron melting are both endothermic processes that require energy and are associated with expansion of the body.  The temperature interval for which melting occurs depends on the composition of the planetesimals and in particular on the amount of water in the anhydrous mineral phase. Iron and silicate melting are also related to porous flow, differentiation of silicates and iron, as well as the associated heat transport. 

\citet{Wak11} investigated the occurrence of liquid water in bodies with radii of up to 1\,000\,km that are heated by $^{26\!}$Al. They considered differentiation of ice and rock via Stokes flow, hydration reactions and varied the time of the (instantaneous) formation and the initial temperature. According to their results, icy planetesimals that formed 2.4\,Ma after CAI will not experience melting of water ice and thus differentiation, regardless of their size. It should be noted, however, that this conclusion must likely be revised toward later formation times if long-lived nuclides and $^{60}$Fe are taken into account \citep[e.g., ][]{McC05}. The thermal history of an icy body --- mostly depending on the formation time relative to the CAIs --- and depending on which of the above processes actually occurred, may result in a variety of interior structures. Figure \ref{GTBSFigstructure} shows some possibilities ranging from a homogeneous body made of a mixture of H$_2$O and high-density silicates to a fully differentiated structure with even an inner iron core. Whether the latter structure is possible, however, is questionable for small icy bodies because large amounts of $^{26\!}$Al or early accretion may rather result in a dehydrated planetesimal as water may vaporize and escape the accreting body \citep{McC05}.

Among the icy bodies, Ceres, the second target of the Dawn spacecraft, is of particular interest and its thermal evolution has been repeatedly modeled. Ceres is the largest body in the asteroid belt with a radius of 470\,km and its low bulk density of $2\,077\pm36$\,kg\,m$^{-3}$ \citep{Tho05} suggests a global ice mass fraction of 17 -- 27 weight-percent if the average porosity is negligible \citep{McC05}. The surface material displays spectral properties of carbonates, brucite and carbon-rich clays \citep{Ray03,Coh98,Riv06,Mil09}. Based on the inferred surface composition, the primordial material of Ceres is of carbonaceous chondritic nature and has been formed during the first 10 million years of the solar system. Taken together the spectral data and mapping, Ceres is most probably an ice-rich world whose surface has substantially been changed by aqueous alteration. The latter could be a result of the eruptions driven by interior aqueous activity and subsequent sublimation of ice from the surface.  Assuming an ice-silicate composition, thermal models by \citet{McC05} which considered short- and long-lived radioactive nuclide heating suggest that Ceres is likely differentiated into a rocky core, an icy mantle and possibly a shallow still liquid layer above the core (Fig.~\ref{GTBSFigstructure}b,c). Even if only long-lived radionuclide heating is assumed, the water ice in Ceres would melt quickly and a water mantle would form, except for an upper crust that would not melt. Whether this structure actually formed depends critically on the amount of water at the time of its accretion, and the amount of $^{26\!}$Al present in the pre-Ceres objects.  The likely presence of a liquid layer above the silicate core has been supported by \citet{Cas10} who emphasize that the surface temperatures of 180\,K would increase the likelihood of a deep ocean in today's Ceres, provided a sufficient concentration of ammonium and salts (Fig.~\ref{GTBSFigstructure}c) are present.

\bigskip
\centerline{\textbf{\refstepcounter{section}\thesection. THERMOCHRONOMETRY OF METEORITES
\label{GTBSSectThermChron}
}}
\bigskip

\begin{table*}[t]

\centerline{\sc Table \refstepcounter{table}\thetable\label{GTBSTabMetDat}}
\smallskip
\centering{\sc Closure ages determined for selected H chondrites for a number of  thermochronometers. Only meteorites are listed for which at least three data points are available. (All temperatures in Kelvin, all ages in Myear.)
}
\\[2.2ex]
{\small
\begin{tabular}{lcr@{}lr@{}lr@{}lr@{}lr@{}lr@{}l}
\hline
\hline
\noalign{\smallskip}
&& \multicolumn{12}{c}{Thermochronometer} \\
 & & \multicolumn{2}{c}{Hf-W} & \multicolumn{2}{c}{Pb-Pb} & \multicolumn{2}{c}{Al-Mg} & \multicolumn{2}{c}{U-Pb-Pb} & \multicolumn{2}{c}{Ar-Ar} & \multicolumn{2}{c}{fission-tracks}\\
\noalign{\smallskip}
\hline
\\[-0.7ex]
\multicolumn{2}{l}{closure temperature}& 1\,150&$\,\pm75\,$ & 1\,050&$\,\pm\,100$ & 750&$\,\pm\,130$ & 720&$\,\pm\,50$ & 550&$\,\pm\,20$ & 390&$\,\pm\,25$ \\
\noalign{\smallskip}
\hline
\noalign{\smallskip}
Meteorite & type & \multicolumn{12}{c}{Closure time} \\
\noalign{\smallskip}
Estacado        & H6 & 4\,557.3&$\,\pm\,1.6$ & 4\,561  &$\,\pm\,7$   &         &         & 4\,501.6&$\,\pm\,2.2$ & 4\,465&$\,\pm\,5$ & 4\,401&$\,\pm\,10$ \\
Guare\~na        & H6 &        &         &         &         &         &         & 4\,504.4&$\,\pm\,0.5$ & 4\,458&$\,\pm\,10$ & 4\,402&$\,\pm\,14$ \\
Kernouve\'e      & H6 & 4\,557.9&$\,\pm\,1.0$ & 4\,537.0&$\,\pm\,1.1$ &         &         & 4\,522.5&$\,\pm\,2.0$ & 4\,499&$\,\pm\,6$ & 4\,438&$\,\pm\,10$ \\
Mt. Browne      & H6 &         &         & 4\,554.8&$\,\pm\,6.3$ &         &         & 4\,543  &$\,\pm\,27$ & 4\,516&$\,\pm\,5$ & 4\,471&$\,\pm\,13$  \\
Richardton      & H5 & 4\,561.6&$\,\pm\,0.8$ & 4\,562.7&$\,\pm\,1.7$ &         &         & 4\,551.4&$\,\pm\,0.6$ & 4\,525&$\,\pm\,11$ & 4\,469&$\,\pm\,14$ \\
Allehan         & H5 &         &         &         &         &         &         & 4\,550.2&$\,\pm\,0.7$ & 4\,541&$\,\pm\,11$ & 4\,490&$\,\pm\,14$ \\
Nadiabondi      & H5 &         &         & 4\,558.9&$\,\pm\,2.3$ &         &         & 4\,555.6&$\,\pm\,3.4$ & 4\,535&$\,\pm\,10$ & 4\,543&$\,\pm\,20$ \\
Forest Vale     & H4 &         &         &         &         & 4\,562.5&$\,\pm\,0.2$ & 4\,560.9&$\,\pm\,0.7$ & 4\,551&$\,\pm\,8$  & 4\,544&$\,\pm\,14$ \\
Ste. Marguerite & H4 &         &         &         &         & 4\,563.1&$\,\pm\,0.2$ & 4\,562.7&$\,\pm\,0.6$ & 4\,562&$\,\pm\,16$ & 4\,550&$\,\pm\,17$ \\
\noalign{\smallskip}
\hline
\noalign{\smallskip}
\multicolumn{2}{l}{Sources of data} & \multicolumn{2}{c}{(a)} & \multicolumn{2}{c}{(b)} & \multicolumn{2}{c}{(c)} & \multicolumn{2}{c}{(d)} & \multicolumn{2}{c}{(e)} & \multicolumn{2}{c}{(f)}  \\
\noalign{\smallskip}
\hline
\end{tabular}
}

\renewcommand{\baselinestretch}{1.0}\scriptsize
\bigskip\noindent
\parbox{0.90\hsize}{\small 
 Keys for sources: 
\\
(a)~Hf-W ages are from \citet{Kle08} and were re-calculated relative to the $^{182}$Hf/$^{180}$Hf of the angrite D'Orbigny, which has a Pb-Pb age of $t = 4\,563.4\pm±0.3$\,Ma \citet{Kle12}. Closure temperature calculated using lattice strain models.
\\
(b)~Closure temperature for Pb diffusion in chondrule pyroxenes estimated by \citet{Ame05}. Age data by \citet{Bli07} for Estacado and Mt. Browne, \citet{Ame05} for Richardton, and \citet{Bou07} for Nadiabondi and Kernouv\'e.
\\
(c)~Data from \citet{Zin02}. Activation energy and frequency factor by \citet{LaT98}: $E = 274$ kJ/mol, $D_0 = 1.2\times10^6$ . Closure temperature calculated according to \citet{Dod73} at 1\,000\,K/Ma cooling rate and $2\,\mu$m feldspar grain size.
\\
(d)~Closure temperature by \citet{Cer91}, phosphate U-Pb-Pb age data by \citet{Goe94}, and \citet{Bli07} for Estacado and Mt. Browne. 
\\
(e)~Ar-Ar ages by \citet{Tri03} and \citet{Sch06} for Mt. Browne, recalculated for miscalibration of K decay constant \citep[, see text]{Ren11,Schw11,Schw12}. Closure temperature by \citet{Tri03} and \citet{Pel97}.
\\
(f)~Calculated age at 390\,K from time interval between Pu-fission track retention in merrillite at 390\,K and Pu-fission track retention by pyroxene at 550\,K
(corresponds to Ar-Ar feldspar age at 550\,K). Data by \citet{Tri03}, closure temperature by \citet{Pel97}.
}

\end{table*}

The physical properties (e.g., size, formation time, short-lived nuclide concentration, initial porosity, heat conduction) of meteorite parent bodies significantly influence their cooling history. Isotopic dating of meteorites can define points in time when cooling caused temperatures to fall below a certain critical value called ``closure temperature".

The age of a mineral or rock, usually expressed in million years (Ma) or billion years (Ga), can be understood as the accumulation time of a daughter isotope by radioactive decay from a parent nuclide. The system to be dated must meet certain conditions to obtain geologically meaningful ages:

1. "Closed system" condition: During the accumulation time no loss, gain --- except radioactive decay --- or diffusion processes affecting concentration profiles of both parent and daughter nuclides should occur, unless such changes can be corrected for.

2. During formation of the system, parent and daughter elements were fractionated, i.e., partitioned differently into different phases due to their different geochemical behaviour. On the other hand, daughter element isotopes were equilibrated, i.e., all cogenetic phases have identical daughter isotope ratios. This allows the determination of the initial abundance of daughter atoms, to quantify the daughter isotope excess due to in situ radioactive decay.

If these conditions are fulfilled, radioisotope ages date the event of fractionation of parent from daughter element. This may be caused by the crystallization of cogenetic minerals, yielding the crystallization age of a rock. Strictly speaking these ages are cooling ages, as the radiometric clock only starts below the so called closure temperature when diffusion of daughter element isotopes becomes ineffective. Closure temperatures depend on the dating system and the diffusion properties of isotopes in the mineral(s) dated. Particularly low closure temperatures are inherent to isotopic dating systems involving high diffusion coefficients such as noble gases. Cooling ages may significantly postdate the crystallization of a rock, if cooling needed long time intervals and closure temperatures are low.

The following radioisotopic dating systems can be used to infer constraints on a meteorite's cooling history:

\medskip\noindent
1. $^{182}$Hf-$^{182}$W:
\\
This system is based on the decay of the short lived nuclide $^{182}$Hf and dates the separation of metal and silicate during parent body metamorphism or differentiation. For chondrite metamorphic processes, the closure temperature of $1\,150\pm75$\,K for H6 chondrites \citep{Kle08} was inferred by lattice strain models. 

\medskip\noindent
2. U-Pb-Pb:
\\
This system is based on the simultaneous decay of $^{238}$U to $^{206}$Pb and $^{235}$U to $^{207}$Pb. High precision ages are usually obtained by phases rich in U, e.g., phosphates \citep{Goe94,Bli07}. Phosphates date cooling through the closure temperature of 720\,K as inferred from diffusion experiments \citep{Cer01}. Silicates hamper Pb diffusion at higher temperatures, \citet{Ame05} estimated $1\,050\pm100$\,K for pyroxenes of chondrules from chondritic meteorites. Silicate dates \citep{Ame05,Bli07,Bou07} are usually less precise, and are used only, if the maximum temperature to which the meteorite was subjected was comparable or exceeded the closure temperature of $1\,050\pm100$\,K. For mildly metamorphosed meteorites, this is frequently not the case. 

\medskip\noindent
3. $^{26\!}$Al-$^{26}$Mg:
\\
The $^{26\!}$Al decays with a half life of 0.72\,Ma. There are many Al-Mg dates for CAIs and chondrules, and very few for parent body processes. For example, two dates exist of formation and cooling of secondary 
feldspar for the H chondrites Forest Vale and Ste. Marguerite \citep{Zin02}. Diffusion data exist for Ca rich feldspar \citep{LaT98}, from which a closure temperature of $750\pm130$\,K can be calculated for a cooling rate of 1\,000\,K/Ma and 2 micrometer sized feldspar grains, using the \citet{Dod73} formula. 

\medskip\noindent
4. $^{40}$Ar-$^{39}$Ar:
\\
This is a classical technique based on the dual decay of $^{40}$K to $^{40}$Ar and $^{40}$Ca: numerous whole rock Ar-Ar data exist, few data exist on mineral separates, e.g., feldspar that usually dominates whole rock ages of equilibrated ordinary chondrites. Ar diffusion data were used to infer a closure temperature of $550\pm20$\,K \citep{Tri03}. Currently there is a debate of the revision of the K decay constant \citep{Schw11,Ren10}. However, most recent recommendations \citep{Schw12,Ren11} agree that the \citet{Ste77} convention should be revised upwards by 22 -- 28\,Ma at an absolute age of 4\,500\,Ma. 

\medskip\noindent
5. $^{244}$Pu fission tracks:
\\
This method uses tracks from the now extinct nuclide $^{244}$Pu registered by phosphates and adjacent silicates (e.g., orthopyroxene). Due to different retention temperatures (550\,K for  orthopyroxene, 390\,K for merrillite obtained by annealing experiments; \citealt{Pel97}), minerals can record different fission track densities in case of slow cooling, whereas fast cooling results in indistinguishable densities. Due to complex correction procedures --- for irradiation geometry, cosmic ray tracks and cosmic ray spallation recoil tracks –-- this method is very time consumptive which explains that only few groups worldwide applied this technique \citep{Pel97,Tri03}. 

\bigskip\noindent
6. $^{53}$Mn-$^{53}$Cr:
\\
For this system, isotopic ages can be obtianed on certain mineral phases as well as acid leachates. The closure temperature for orthopyroxene and olivine for typically 100 micrometer sized grains is around 800 -- 900\,K \citep{Ito06,Gan07}.

\medskip\noindent
7. $^{129}$I-$^{129}$Xe:
\\
The $^{129}$I-$^{129}$Xe system can be applied to ordinary chondrites phosphate or feldspars as well as hydrous phases from carbonacous chondrites \citep{Bra99}. The closure temperature of the I-Xe system is largely unconstrained, however, the agreement of some Pb-Pb and I-Xe phosphate data of slowly cooled ordinary chondrites suggest a similar closure temperature of 720\,K of the I-Xe system in chondritic phosphates. 

\medskip\noindent
In order to derive meaningful constraints on the parent body thermal evolution, it is necessary to ascertain that the thermochronological data are due to the primordial cooling history of the parent body, and not due to secondary impact induced reheating or other shock effects. As meteorites are derived from asteroids that were fragmented by collisions occurring throughout the 4.6\,Ga history of the solar system, meteorite samples that display shock metamorphism and implicitly experienced disturbances of radioisotopic clocks should be avoided. In practice that means that only meteorites displaying a low degree of shock metamorphism should be selected for thermochronological evaluation. 

Apart from that, reasonable thermochronological constraints can only be obtained for meteorites for which several points on a cooling curve exist, i.e., different radioisotopic dating techniques should have been applied that define the points of time at which temperatures fell below different --- system specific --- closure temperatures. The biggest such data set exists for H chondrites; it is collected in Table~\ref{GTBSTabMetDat}.

\bigskip
\centerline{\textbf{\refstepcounter{section}\thesection. MODEL CALCULATIONS FOR CHONDRITIC\label{GTBSSectModels}
}}
\centerline{\textbf{PLANETESIMALS
}}
\bigskip

\noindent
\textbf{\refstepcounter{subsection}\thesubsection\ Structure of chondritic material}
\bigskip

Detailed studies of the thermal evolution of planetesimals with particular emphasis on comparison with the meteoritic record are presently only available for parent bodies of ordinary chondrites, hence we concentrate on such objects. The main constituents present in chondritic meteorites are chondrules and a matrix of fine dust grains. The typical relative abundances of these constituents are shown in Table~\ref{GTBSTabMetTypes}, they have typical sizes of micrometers in the case of relatively unprocessed matrix \citep{Sco04}, and typical sizes of millimetres in  the case of chondrules, with variations specific to each chondrite group \citep{Sco97,Weis06}. Though chondrules may also contain fine grained mineral assemblages, they entered the meteorite parent body as solidified melt droplets after they formed by unspecified flash heating events in the solar nebula.

The compacted solid material has a density $\varrho_{\rm b}$. This material is a complex mixture of minerals, dominated by olivine and pyroxenes. The porous material has a density $\varrho=\varrho_{\rm b}(1-\phi)+\varrho_{\rm p}\phi$, where $\varrho_{\rm p}$ is the density of the material in the pores. For ordinary chondrites the pores are filled with a low pressure gas and this may be neglected, for carbonaceous chondrites the pores could be filled by water. Besides porosity $\phi$ one also considers the filling factor $D=1-\phi$.

The porous material may approximately be described as a packing of spheres. For a packing of equal sized spheres it is found that there are two critical filling factors. One is the random close packing with a filling factor of $D=0.64$, i.e., a porosity of $\phi=0.36$ \citep{Sco62,Jae92,Son08}. The other one is the loosest close packing that is just stable under the application of external forces (in the limit of vanishing force), which has $D=0.56$ or $\phi=0.44$ \citep{Ono90,Jae92}.

At the basis of the planetesimal formation process there stands the agglomeration of fine dust grains like that found in IDPs into dust aggregates of increasing size. Such an agglomerated material from very fine grains that was not subject to any pressure has porosity as high as $\phi\approx0.8\dots0.9$ \citep[e.g.][]{Yan00,Kra11}. Such high-porosity material seems to be preserved in some comets \citep{Blu06}. 

Collisions of dust aggregates  during the growth process of planetesimals leads to compaction of the material. The experiments of \citet{Wei09} have shown that the porosity can be reduced to $\phi\approx 0.64$ \citep[or even lower, see][]{Kot10} by repeated impacts. This porosity is still higher than the random loose packing, lower porosities of $\phi \lesssim0.4$ were obtained by applying static pressures of more than 10\,bar \citep{Gue09}. Since planetesimals form by repeated collisions with other planetesimals and impact velocities can be rather high (see chapters by Johansen et al.) one can assume in model calculations that the initial porosity of the material from which the parent bodies of the asteroids formed is already compacted to~$\phi\approx0.4$.

\bigskip
\noindent
\textbf{\refstepcounter{subsection}\thesubsection\ Equations}
\bigskip

Essentially all the models for the internal constitution and the thermal evolution of the parent bodies of meteorites considered so far are one dimensional spherically symmetric models. The basic equations to be considered in this case are the heat conduction equation 
\begin{equation}
\frac{\partial T}{\partial t} = \frac{1}{\varrho c_p r^2} \frac{\partial}{\partial r} r^2 K \left(T,\phi \right) \frac{\partial T}{\partial r} + \frac{H}{c_p}\,,
\label{GTBSEqHeat}
\end{equation}
where $T$ is the temperature, $c_p$ the specific heat per unit mass of the material, $K$ the heat conductivity, $H$ the heating rate per unit mass, and $\varrho$ the mass-density, and the  hydrostatic pressure equation
\begin{equation}
\frac{\partial P}{\partial r}=\frac{G M_r}{r^2} \cdot \varrho\quad\mbox{with}\quad M_r=\int_0^r4\pi\varrho(s)s^2\,{\rm d}s\,,
\label{GTBSEqHydro}
\end{equation}
where $G$ is the gravitational constant and $M_r$ is the mass inside a sphere of radius $r$. They have to be combined with an equation of state $\varrho=\varrho\left(P,T,D\right)$ and an equation for the compaction of the chondritic material which describes how the filling factor $D$ of the granular material develops over time under the action of pressure and temperature. The equations have to solved subject to the appropriate initial and boundary conditions. 

If compaction and melting of the material is not considered one can assume $\varrho$ to be constant. The pressure equation and the equation of state usually need not to be considered in that case. This is an approach that is frequently followed if only the effects of heating and cooling on the chondrite material are of interest (see Section \ref{GTBSAnalytSol}).

For the initial conditions the question becomes important how the big planetesimals are formed that could be the parent bodies of meteorites. Though the details of this are not undoubtedly known, the general assumption is that they assemble from smaller planetesimals that in turn somehow assembled from the dust in the protoplanetary accretion disk. Essential for the problem of thermal evolution of the bodies is how the time during which the parent body acquired most of its material is related to the duration of the main heating processes acting in the body. If decay of short lived radioactive nuclei are the main heat source, the body is supplied with the whole latent heat content of the radioactive nuclei if the essential growth period is short compared to the decay constant $\lambda$ of the dominating heat source, while in the opposite case much of the latent heat present at the onset of growth is liberated and radiated away before the material is added to the growing body.

The case that the body forms rapid compared to the timescale $\lambda$ is denoted as \emph{instantaneous formation}. In case of slow formation one has to add to the above equations a prescription how the mass $M$ of the body increases during the growth process. In principle the mass of the big planetesimals increases stepwise during run-away growth by the rare events of impacts with the most massive bodies from the background planetesimal swarm. This growth mode was studied, e.g., by \citep{Gho03}. Due to the difficulties in modelling unsteady growth, most of the few studies considering growth assume a continuous growth with some prescribed rate $\dot M(t)$ of mass-increase. In such calculations one typically starts with some small seed body and follows its growth as matter is deposited at the surface.

The temperature at the surface, $T_{\rm srf}$, is determined by the gain and loss of energy at the surface in form of a boundary condition at radius $R$
\begin{equation}
\left.-K\frac{\partial T}{\partial r}\right\vert_{r=R}= H_{\rm srf}-L\,,
\end{equation}
where the left hand side is the heat flux from the interior to the surface, $H_\mathrm{srf}$ is the energy gain from the outside and $L$ is the energy lost to exterior space. The most important processes to be considered are heat gain by absorption of radiative energy from outside and heat loss by radiation into empty space ($\propto T^4_{\rm srf}$). For growing bodies also the heat content of the infalling material and impact heating have to be considered, but for bodies of less that about 500\,km size impact heating can be neglected. (To be more accurate: Impact heating is important as a local effect, but not for the global energy budget, e.g., \citealt{Kei97,Sra12}.)  

Since the heating by irradiation by the sun (or by the accretion disc during the earliest phases) varies with distance, it is practically not feasible to determine  $T_{\rm srf}$ from this boundary condition because the location of a particular parent body of meteorites at its formation time is not known. Therefore one has to introduce some ad-hoc prescription for $T_{\rm srf}$; typically one prescribes some representative value for the surface temperature of bodies in the asteroid belt and assumes that this temperature is constant over the first few hundred million years of thermal evolution.

\begin{figure}[t]
\epsscale{1.0}

\plotone{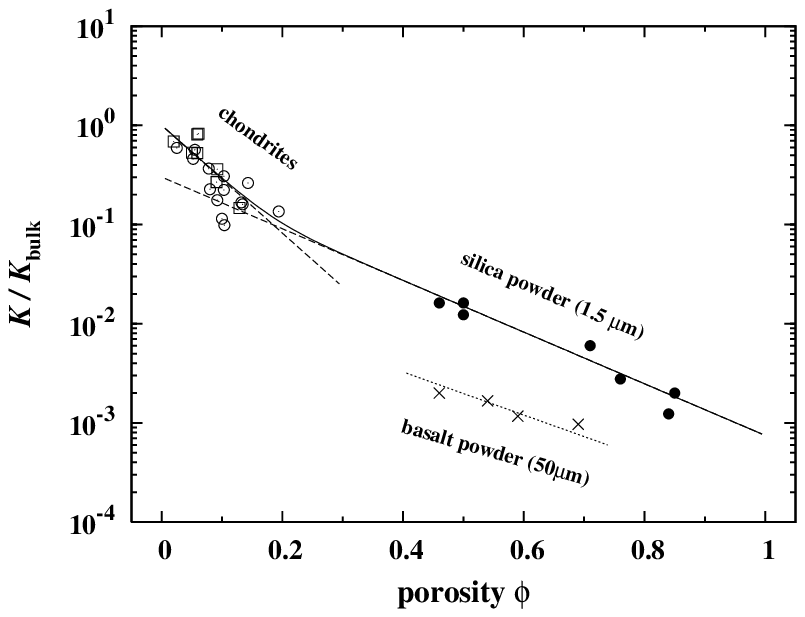}

\caption{\small Variation of heat conductivity $K$ with porosity $\phi$. Results for fine grained silica powder (filled circles) from experiments of  \citet{Kra11,Kra11b}, and for particulate basalt (crosses) from \citet{Fou70}. Typical grain size is indicated for both cases. Open squares and open circles are experimental results for heat conductivity and porosity for ordinary H and L chondrites, respectively,  from \citep{Yom83} and the solid line is the fit according to Eq.~(\ref{GTBSFitToKpor}) \citep[from][]{Hen12a}.}

\label{GTBSFigKpor}
\end{figure}

\bigskip
\noindent
\textbf{\refstepcounter{subsection}\thesubsection\ Heat capacity and heat conductivity\label{GTBSProMod}}
\bigskip

The heat conduction equation depends on the specific heat $c_p$ per unit mass. This can be calculated from the values of $c_p$ for the components of the mineral mixture. For the material of H and L chondrites \citet{Yom84} give an analytic fit for the temperature variation of $c_p$ that can be used in model calculations. 
In many model calculations a fixed value for $c_p$ corresponding to room temperature is used, but this significantly underestimates the real $c_p$ at higher temperatures and results in too high temperatures. The importance of using realistic temperature dependent specific heats for modelling the thermal evolution of planetesimals has been stressed by \citet{Gho99}. 

A good knowledge of the heat conduction $K$ coefficient is particular important for calculating reliable thermal models. Unfortunately the heat conduction strongly depends on the structure of the material and no simple method is known to calculate this for the highly complex structured material of chondrites. \citet{Yom83} measured the heat conductivity of chondritic material for a sample of H and L chondrites in the temperature range 200 -- 500\,K. The result shows a moderate temperature variation and a strong dependence on the porosity. They developed a fit-formula \citep{Yom84} that describes the temperature-porosity  variation of an average heat conductivity $K$ for this  material. A more recent experimental investigation of $K$ for meteoritic material \citep{Ope10} found similar results.

\begin{figure*}[t]
 \epsscale{1.6}

\centerline{\plottwo{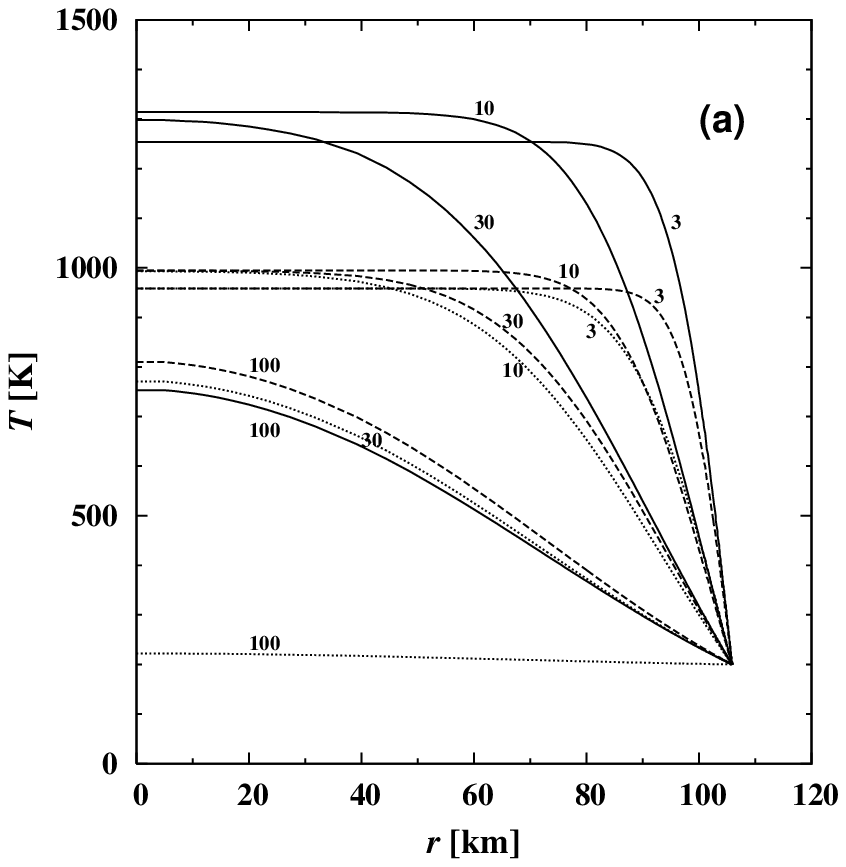}{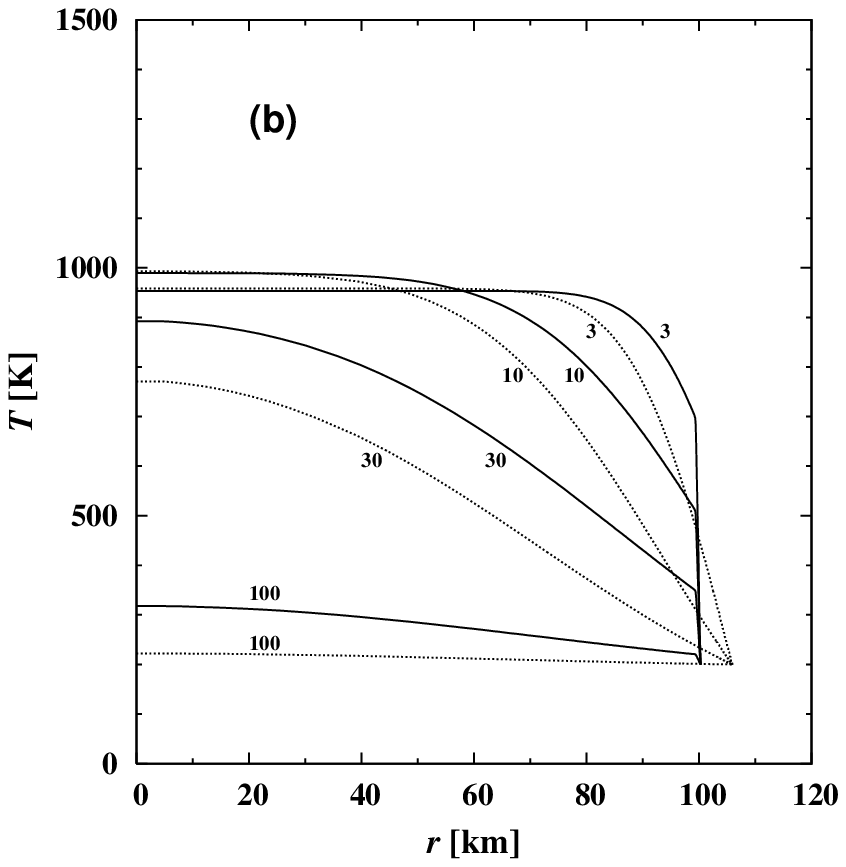}}

\centerline{\plottwo{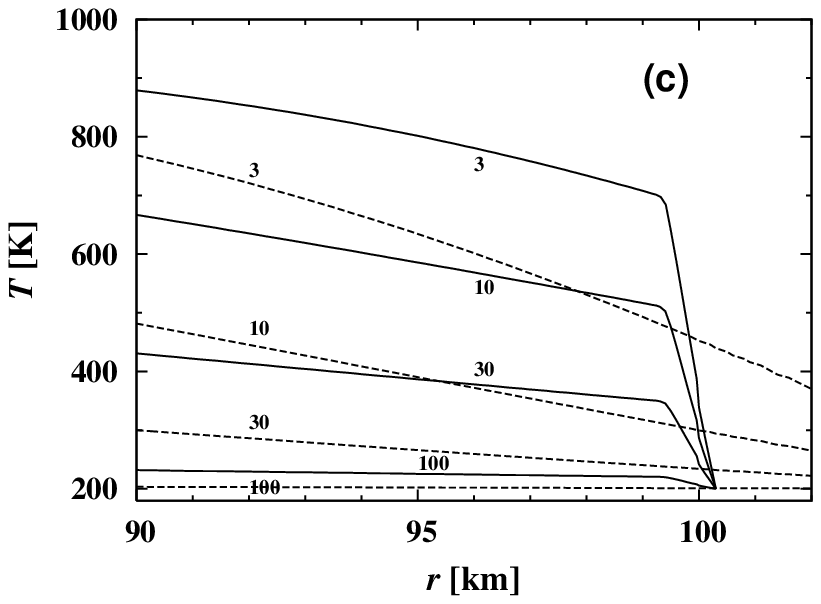}{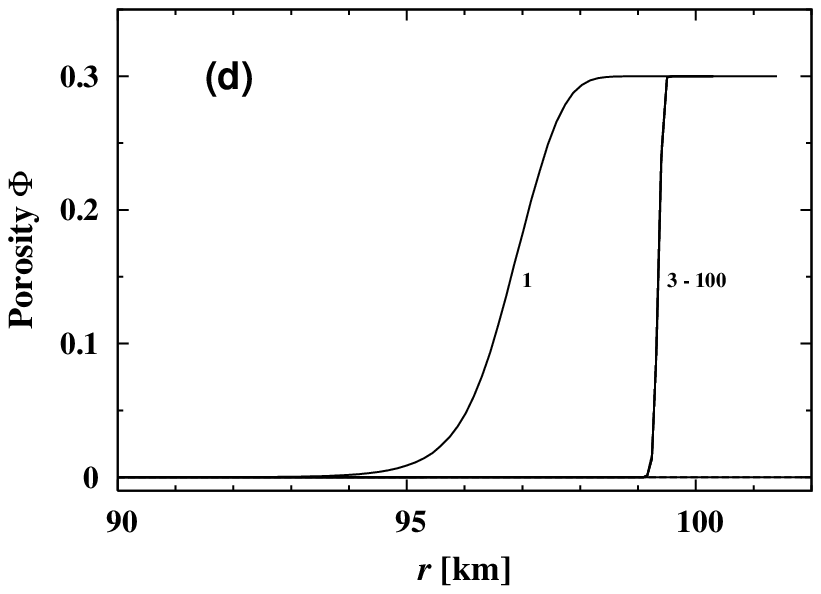}}

\caption{\small Comparison of models for the thermal evolution of planetesimals based on different approximations. The formation time is assumed to be $t_{\rm form}=2.2\times10^6$\,a after CAI formation, instantaneous formation assumed. The surface temperature is fixed at $T_{\rm srf}=200$\,K. The heating is assumed to be due to $^{26\!}$Al only, the rate is $h=1.82\times10^{-7}$\,W/kg at time of CAI formation. The density of the bulk material is $\varrho_{\rm b}=3\,800$ kg/m$^3$; the initial density $\varrho_0$ is lower in some models, corresponding to a certain initial porosity $\Phi$ of the material. The initial radius is determined such that all models have the same mass of $1.6\times10^{19}$\,kg, which corresponds to a body of $R=100$\,km radius if the matter is compacted to $\varrho_{\rm b}$.  Shown are the radial temperature variations at the indicated instances (in Ma after $t_{\rm form}$).
Panel (a): Models without sintering or outer regolith layer. Full line: Model like \citet{Miy81} with $\varrho_0=3\,200$\,kg/m$^{3}$, fixed specific heat capacity $c_p=625$\,J/kgK, fixed heat conductivity $K=1$\,W/mK. Dashed line: Same model, but temperature dependent heat capacity in the analytic approximation given by \citet{Yom83}. Dotted line: Same model, but $K=3.7$\,W/mK, corresponding to almost compacted material. Panel (b): Comparison of models with and without insulating outer layer. Dotted line: Same model as in panel a. Full line: Sintering included \citep[as in][]{Hen12a} with an assumed initial porosity of $\Phi=0.3$. Panel (c): Shows the outermost layers of the models in panel b. Panel (d): Distribution of porosity in the model with sintering. }

\label{GTBSFigCompModels}
\end{figure*}

There is, however, the problem that the measurements are done for material that is already strongly compacted with porosities generally less than 15\% for unshocked material \citep{Bri03,Con08} while the initial porosity of the material must have been much higher than $\approx40$\%. The heat conductivity of highly porous material relevant for the initial dust material from which planetesimals formed has been measured by \citet{Kra11} in the range of porosities from 44\% to 82\%. This has been matched by \citet{Kra11b} and \citet{Hen12a} with the results of \citet{Yom83} into a unified analytic approximation for the whole range of porosities of the form
\begin{equation}
K=K_{\rm b}\,{\cal K}(\phi)\,,
\end{equation}
where $K_{\rm b}$ is the --- possibly temperature dependent --- heat conductivity of the bulk material, and ${\cal K}$ describes the variation with porosity (${\cal K}=1$ for $\phi=0$)
\begin{equation}
{\cal K}(\phi)=\left({\rm e}^{-4\,\phi/\phi_1}+{\rm e}^{4\,(a-\phi/\phi_2)}\right)^{1/4}
\label{GTBSFitToKpor}
\end{equation} 
with constants $a$, $\phi_1$, $\phi_2$ given in its latest form in \citet{Hen13}. An essentially equivalent approach was developed by \citet{War11}. Figure \ref{GTBSFigKpor} illustrates how strongly the heat conductivity changes with porosity because in a granular material heat flows only through the small contact areas of the particles. The heat conductivity changes by two orders of magnitude if the porosity varies from $\phi\approx40$\% from the loosely packed initial stage to $\phi\approx0$ for consolidated material. By this a porous surface layer that escapes compaction completely changes the internal thermal structure by its insulating effect, as is demonstrated in \citet[see also Fig.~\ref{GTBSFigCompModels}]{Akr98}.

The bulk heat conductivity $K_{\rm b}$ of the material is that of a complex mixture of minerals with iron and FeS. There is no exact method to calculate this for composite media, but a number of recipes have been developed for approximate calculations \citep[see][ for a review]{Ber95}.  
 
\bigskip
\noindent
\textbf{\refstepcounter{subsection}\thesubsection\ Heating\label{GTBSSecTHeating}}
\bigskip

The parent bodies of ordinary chondrites where subject to substantial heating because the mineral content of the meteorites of high petrologic types was equilibrated at temperatures up to 1\,000\,K and even above \citep[e.g.][]{Sla05}. The possible heat sources responsible for heating have been debated controversial \citep[see ][ for a review]{McS03}. Presently the most efficient sources are believed to be heating by decay of short lived radioactive nuclei and, for bodies with diameters in excess of about 500\,km, the liberation of gravitational energy during growth of the bodies. The most efficient sources of heat are shown in Table~\ref{GTBSTabExtinctNuc} \citep[see also][] {Coh00}. The dominating heat source is decay of $^{26}$Al, and also $^{60}$Fe was considered, but now its low abundance \citep{Tan12} seems to render its contribution insignificant.

The rate of a radioactive heat source varies with time as
\begin{equation}
H(t)=H_0\,{\rm e}^{-\lambda/t}
\label{GTBSEqHeatVar}
\end{equation}
where
\begin{equation}
\lambda=\ln 2/\tau_{1/2}
\label{GTBSEqDecayConst}
\end{equation}
is the decay constant of the isotope that is responsible for heating and $H_0$ the heating rate by radioactive decay at some initial instant $t=0$. Values for $H_0$ at time of CAI formation and for $\tau_{1/2}$ for the main heat sources are given in Table~\ref{GTBSTabExtinctNuc}). 

\begin{figure}[t]
\epsscale{1.0}

\plotone{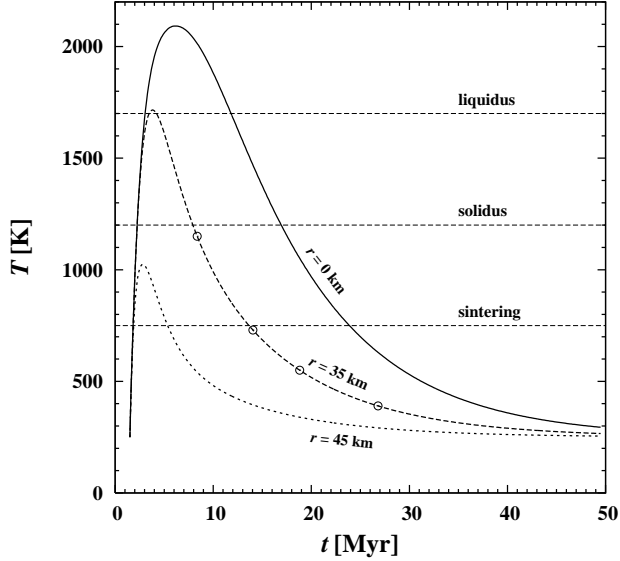}

\caption{\small Thermal evolution of a model asteroid satisfying the prerequisites for applying the analytic solution (\ref{GTBSEqTEvolAstMiy}). The radius $R$ is 50\,km , the formation time is $t_{\rm form}=1.5\,\times10^6$\,a after CAI formation. Shown are the temperature evolution at the centre (solid line), at 15\,km depth below the surface (thick dashed line), and at  5\,km depth (dotted line). The thin horizontal dashed lines show three characteristic temperatures important for the thermal evolution of ordinary chondrite parent bodies: The onset of compaction  by sintering, the onset of melting of the Fe-FeS eutectic (solidus) and the final melting of the silicates (liquidus). The four circles denote closing temperatures of 1\,150, 730, 550, 390\,K for different thermochronological systems, see Table~\ref{GTBSTabMetDat}. The model parameters are chosen as in \citet{Hev06}: $K=2.1$\,W/mK, $c_p=837$\,J/kgK, $\varrho=3\,300$\,kg/m$^3$, $T_0=250$\,K, heating by decay of $^{26\!}$Al with $\lambda=9.5\times10^{-7}$\,a$^{-1}$ and $H_0=1.9\times10^{-7}$\,W/kg.}
\label{GTBSFigTherEvoExpl}
\end{figure}

\bigskip
\noindent
\textbf{\refstepcounter{subsection}\thesubsection\ Analytic model\label{GTBSAnalytSol}
}
\bigskip

The most simple case that could be considered is if the heat capacity $c_p$, heat conductivity $K$, and matter density $\varrho$ all are assumed to be constant and if the body is formed instantaneously at some instant taken as the origin of time-axis. Only the heat transport equation~(\ref{GTBSEqHeat}) has to be solved in this case. The initial-boundary value problem with constant temperature $T_0$ at initial time $t=0$ and at the fixed outer boundary at $r=R$, and a heat source varying with time as in Eq.~(\ref{GTBSEqHeatVar}), allows for an analytic solution \citep{Car59} 
\begin{eqnarray}
T(r,t)&=&T_0+{H_0\over c_p\lambda}\,{\rm e}^{-\lambda t}\,\left[{R\sin r\left(\lambda\over\kappa\right)^{1\over2}\over r\sin R\left(\lambda\over\kappa\right)^{1\over2}}-1
\right] \nonumber
\\[.5cm]
&&\hskip-1.7cm
+{2R^3 H_0\varrho\over r\pi^3 K}
\ \sum_{n=1}^\infty\ 
{\displaystyle(-1)^n\,\sin{n\pi r\over R}
 \over \displaystyle
 n\left(n^2-{\lambda R^2\over\kappa\pi^2}
\right)
}\ 
{\rm e}^{-\kappa n^2\pi^2t/R^2}\,,
\label{GTBSEqTEvolAstMiy}
\end{eqnarray}
where $\kappa=K/\varrho c_p$ is the heat diffusivity. If more than one radioactive heat source is present, the second and third term of the right hand side have to be repeated for each decay system. Though this solution neglects important aspects (significant spatial and temporal variations of $K$, $c_p$, $\varrho$, finite formation time and so on), it was and still is used in many studies of thermal evolution of asteroids because it enables one to study qualitatively some basic features of the problem in a simple way.

Figure \ref{GTBSFigTherEvoExpl} shows as example the thermal evolution of a body of 50\,km size calculated with parameters as used by  \citet{Hev06} who studied the evolution of such a body through the stages of sintering and melting. The figure shows the critical temperatures of $T\approx750$\,K for the onset of sintering of the granular material (see Section~\ref{GTBSSectSinter}), $T\approx1\,200$\,K for the onset of the melting of the Fe-FeS eutectic \citep[see][ for a discussion of the melting behaviour of the Fe-FeS system]{Fei97}, and $T\approx1\,700$\,K where the last silicate components of the chondritic material melt \citep[see][for a study of the melting behaviour of the Allende chondrite]{Age95}. 

\bigskip
\noindent
\textbf{\refstepcounter{subsection}\thesubsection\ Compaction\label{GTBSSectSinter}
}
\bigskip

The compaction of material in planetesimals is a two-step process.  The initially very loosely packed dust material in the planetesimals comes under increasing pressure by the growing self-gravity of the growing bodies. The granular material can adjust by mutual gliding and rolling of the granular components to the exerted force and evolves into configurations with closer packing. The ongoing collisions with other bodies during the growth process enhances this kind of compaction of the material. This mode of compaction, ``cold pressing'', by its very nature does not depend on temperature and operates already at low temperature. This is discussed in \citet{Gue09} and applied to planetesimals in \citet{Hen12a}. 

A second mode of compaction commences if radioactive decays heat the planetesimal material to such an extent, that creep processes are thermally activated in the lattice of the solid material. The granular components then are plastically deformed under pressure and voids are gradually closed. This kind of compaction by ``hot pressing'' or `sintering'' is what obviously operated in ordinary chondrite material, and the different petrologic types 4 to 6 of chondrites are obviously different stages of compaction by hot pressing. \citet{Yom84} where the first to perform a quantitative study of this process by applying the theory of \citet{Rao72}. The basic assumption in  this and all modern theories is that the strain rate $\dot\epsilon$ in the material is related to the applied stress $\sigma$ and the size $d$ of the granular units by a relation of the form
\begin{equation}
\dot\epsilon=A \cdot \frac{\sigma_1^{n}}{d^m} e^{E_{\rm act}/RT}\,,
\label{GTBSDefPowCreep}
\end{equation}
with some constants $A$, $n$, $m$ and activation energy $E_\mathrm{act}$ \citep[see][ for a review on the deformation behaviour of minerals]{Kar13}, and that $\dot\epsilon$ is given in terms of the rate of change of the filling factor as
\begin{equation}
\dot\epsilon=-\dot D/D\,.
\label{GTBSRateStrain}
\end{equation}
The stress $\sigma_1$ is the pressure acting at the contact faces of the granular units. The quantities $A$, ...\ have to be determined experimentally for each material \citep[see][ for a compilation of data]{Kar13}. 

The stress $\sigma_1$ is given by the pressure acting at the contact areas between the granular units. It is assumed that this is given in terms of the applied pressure $P$ and the areas of contact faces, $\pi a^ 2$, and  average cross-section of the cell occupied by one granular unit, $C_{\rm av}$, as $\pi a^2\sigma_1=C_{\rm av}P$. The second basic assumption in the theory refers to the relation between the filling factor $D$, the radius $a$ of the contact areas, the average number of contact points $Z$, and the average cross-section $C_{\rm av}$. Complete sets of equations for calculating this from the geometry of packing are given in \citet{Kak67,Rao72,Arz83,Yom84,Hel85}. In conjunction with these equations, Eq.~(\ref{GTBSRateStrain}) forms a differential equation for calculating the time evolution of $D$ that has to be solved together with the heat conduction and pressure equation. 

Calculations for sintering of the material in planetesimals on this basis have been performed by \citet{Yom84, Akr97, Sen04, Hen12a}. The results suggest sintering in planetesimals of the 100\,km size-class to occur around 750\,K if coefficients in Eq. (\ref{GTBSDefPowCreep}) are taken for olivine from \citet{Sch78}. The use of these coefficients at a much low temperature than for which they have been measured (typically in the range 1\,300 to 1\,600\,K) bears the risk, however, for a loss of reliability of the model results, but no better information is presently available.

\begin{figure}[t]
\epsscale{0.6}

\vskip .4cm
\centerline{
\plotone{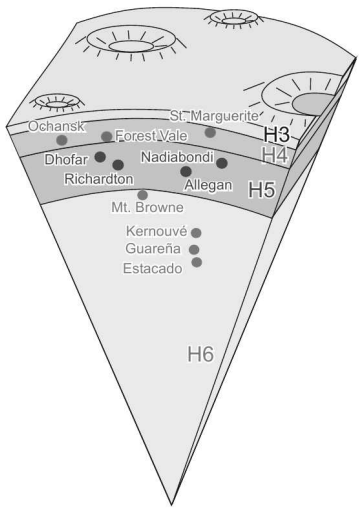}
}

\caption{\small Schematic representation of the onion shell model and the burial depths of the set of H chondrites used for reconstructing the parent body within from a thermal history model of the body and the individual meteorites.}

\label{GTBSFig6Hebe}
\end{figure}

\bigskip
\centerline{\textbf{\refstepcounter{section}\thesection. RECONSTRUCTION OF PARENT BODY\label{GTBSSectReconstr}
}}
\centerline{\textbf{PROPERTIES
}}

\bigskip

\bigskip
\noindent
\textbf{\refstepcounter{subsection}\thesubsection\ Onion shell model}
\bigskip

In the onion shell model it is assumed that the different petrologic types 3 to 6 of meteorites result from different evolution histories of the material under the different temperature and pressure conditions encountered at their burial depth within their parent bodies, see Fig.~\ref{GTBSFig6Hebe}. If the onion shell model is accepted, it is reasonable to assume that meteorites of different petrologic types, if they can be assumed for some reason to originate from the same parent body, carry information on the properties and thermal history of this parent body and on the metamorphic processes within the body in general, and on the processes acting at the initial burial depths of each of the meteorites in particular. This allows one to retrieve the major properties of the corresponding  parent body and its early evolutionary history by comparing empirically determined thermal evolution histories of individual meteorites with thermal evolution models of asteroids. 

The thermal history of a meteorite is derived from the closure ages and the corresponding closure temperatures of a number of radioactive decay systems, which allow us to determine the instant at which the temperature of a meteorite cooled below the closure temperature for diffusion of the decay products of an isotope out of its host mineral (section~\ref{GTBSSectThermChron}). If such kind of data are available for a meteorite for more than one decay system one can try to reconstruct the properties of the parent body and the burial depth of the meteorite within its parent body because the details of the temperature evolution at some location within the body shows characteristic differences for different locations and different bodies. Because of the significant errors of the empirical data this is hardly applicable, however, to a single meteorite (though not completely impossible if at least four data points are available). If, on the other hand side, there is a group of meteorites of different petrologic types with likely origin from one parent body, it is possible to determine the characteristic parameters of the parent body (e.g., size, formation time)  and the individual burial depths of the meteorites, by minimizing the difference between  observed closure ages and temperatures and individual temperature histories for the meteorites at their burial depths within the unknown parent body. 


A good example of a case study describing early heating and cooling of an asteroid could be performed with the H chondrite parent body. There exists a not so small set of H chondrites, for which a number of relatively precise radioisotopic ages is available with different closure temperatures. This data set comprises nine H chondrites of petrologic types 4 -- 6. Table~\ref{GTBSTabMetDat} is a collection of data available for such studies. A lot of these data have only become available during about the last decade such that only now a sufficient data basis exist for such kind of work. 

\begin{table*}[t]

\centerline{\sc Table \refstepcounter{table}\thetable\label{GTBSTabBodyParm}
}
\smallskip
\centering{\sc Properties of the parent body of H chondrites derived by different thermal evolution models, and some key data used in the modelling.}
\\[2.2ex]
{\small
\begin{tabular}{llllllllll}
\hline
\hline
\noalign{\smallskip}
 & \multicolumn{8}{c}{Models}\\
\noalign{\smallskip}
Quantity & Miy81 & Ben96 & Akr98 & Tri03 & Kle08 & Har10 & Hen12 & Mon13 & Unit\\
\noalign{\smallskip}
\hline
\\[-0.7ex]
$t_{\rm form}$   & 2.5 & 2.2 &  2.3  &  2.5 & 2.7 & 2.2 &  1.84 & 1.98 & Ma \\
$R_{\rm core}$   & 85  & 99.0 & 97.5 & 100 & 100 & 99.2 & 110.8 & 126 & km\\
$d_{\rm srf}$    & --- & 1.0 &  2.5  & --- & --- &  0.8 &  0.73 & 0.0 & km\\
$\Phi_{\rm srf}$ & --- &     &       & --- & --- &      &  22.0 &     & \% \\
$T_{\rm srf}$    & 200 & 300 &  200  & 300 & 300 & 170  &  178  & 292 & K \\
\noalign{\medskip}
$M$              & $0.82$ & na & $1.4$ & $1.3$ & $1.3$ & $1.4$ & $2.2$ & 3.2 & $10^{19}$\,kg \\
$K$              & 1 & (a) & (a) & 1 & 1 & (a) & 1.75 (b) & (c) & W/mK \\
$c_p$            & 625 & (d) & (e) & 625 & 625 & (d) & (d) & (e) & J/kgK \\
$\varrho_{\rm bulk}$ & 3\,200 & 3\,780 & 3\,450 & 3\,200 & 3\,200 & 3\,250 & 3\,780 & 3\,800 & kg/m$^3$ \\
$h$              & $1.82$ & na &  & $1.82$ & $1.82$ & $1.82$ & $1.87$ & 1.93 & $10^{-7}$\,W/kg \\
\noalign{\medskip}
heating & $^{26\!}$Al & $^{26\!}$Al &  $^{26\!}$Al, $^{60}$Fe & $^{26\!}$Al & $^{26\!}$Al & $^{26\!}$Al &   $^{26\!}$Al, $^{60}$Fe & $^{26\!}$Al & \\
 & & & l.l. &  & & & l.l. \\
\noalign{\medskip}
$\displaystyle\mbox{\small Solution}\atop\displaystyle\mbox{\small method\ }$\hfill  & analytic & analytic & numeric & analytic & analytic & numeric  &
numeric &  numeric & \\
\noalign{\smallskip}
Sintering & --- & yes & no  & --- & --- & no & yes & no & \\
Regolith  & --- & yes & yes & --- & --- & yes & no & yes & \\
\noalign{\smallskip}
\hline
\end{tabular}
}

\renewcommand{\baselinestretch}{1.0}\scriptsize
\bigskip\noindent
\parbox{.9\hsize}{\small 
Keys for models:
\\
Miy81: \citet{Miy81}, Be96: \citet{Ben96}, Ak98: \citet{Akr98}, Tr03: \citet{Tri03}, Kl08: \citet{Kle08}, Ha10: \citet{Har10}, He12: \citet{Hen12b}, Table 6.
\\[.3ex]
Notes:
\\
(a)~Fit of $K$ to meteoritic data \citep{Yom84}. At $T = 500$\,K  the value of the consolidated material is 3.66W/mK. It  varies only slightly with temperature. 
\\
(b)~Variation of $K$ with prosity according to Eq.~(\ref{GTBSFitToKpor})
\\
(c)~Temperature dependence $K=K_0(T_{\rm srf}/T)^a$ with $K=4$\,W/mK and $a=0,\dots2$, prefered value $a=1/2$.
\\
(d)~Fit of $c_p$  to meteoritic data \citep{Yom84}. At $T = 500$\,K  the value of the consolidated material is 865\,J/kgK.
\\
(e)~Akr98: Data for forsterite, Mon13: data for olivine with admixture of iron.
\\[.3ex]
Abbreviations:
\\
na:~No data given
\\
l.l.:~Contribution of $^{40}$K, Th, U considered}

\end{table*}

Unfortunately, for no other meteorite class there exist a dataset of a comparable size and quality as for H chondrites that can be used for promising attempts to reconstruct its parent body, though there have been a few thermochronological studies for other classes. In the case of L chondrites there are, in fact, also many thermochronological data available, but the parent body probably suffered a catastrophic collision 500\,Ma ago \citep{Kor07} and all thermochronometers are likely to be shifted in this case. For this reason we restrict our discussions to the case of H chondrites. 

Though well defined complete age data sets are rare, recent studies yielded important progress for Mn-Cr dating of aqueous alteration or metamorphism  of the carbonaceous chondrites \citep{Fuj12,Doy13}. \citep{Fuj12} found indistinguishable Mn-Cr ages implying CM carbonate formation $4.8\pm0.5$\,Ma after CAIs, their parent body modeling yielded parent body accretion 3.5\,Ma after CAIs. This study also solved a critical problem in carbonate Mn-Cr dating, i.e., evaluation of precise dates by establishing correct sensitivity coefficients of Mn and Cr in carbonates, invalidating some previously published Mn-Cr age data.

The onion shell modell is criticised heavily by some authors \citep[e.g.][]{Tay87,Sco81,Gan13,Sco13}, because metallographic cooling data are argued to suggest an impact-dominated thermal evolution. Though such problems demonstrate that the thermal evolution does not proceed fully in such a simple way as it is assumed in the onion-shell model,  the models to be discussed later show, on the other hand, that it is possible to obtain a fully consistent picture for the thermal history of the H chondrite parent body from the onion-shell model. The solution for this contradictory situation remains to be awaited. 

\bigskip
\noindent
\textbf{\refstepcounter{subsection}\thesubsection\ Reconstructions of the H chondrite parent body}
\\[.5cm]
There were a number of attempts to reconstruct the properties of the H chondrite parent body using empirically determined cooling histories of H chondrites from different initial burial depths within the parent body. Table \ref{GTBSTabBodyParm} gives an overview on the results, the values of key parameters of these models, and of the assumptions and methods used. The table lists all the models that attempted to obtain in some way a ``best fit'' between properties and cooling histories derived from laboratory studies and a thermal evolution model of the parent body. The results are radius $R$ and formation time $t_{\rm form}$ of the putative parent body, and the burial depths of the meteorites. All studies showed, that the observed properties of the investigated H chondrites are compatible with a parent body of radius of the order of magnitude of 100\,km and an ``onion shell'' model. In this model the body is accreted cold, then heats up within a period of ca. 3\,Ma by decay of radioactive nuclei, mainly $^{26\!}$Al, and finally cools down over an extended period of ca. 100\,Ma, without suffering catastrophic collisions that would have disrupted the body, thereby disturbing its thermal evolution record. All models published so far are based on the instantaneous formation hypothesis; in all cases the body is assumed to have a spherically symmetric structure.

The first such model was constructed by \citet{Min79}. The small number of data available at that time allowed only some very crude estimations of parent body radius and formation time. This first model initiated further studies based on better data and methods. Many of such models are based on the analytic solution of the heat conduction equation (\ref{GTBSEqHeat}) with constant coefficients and fixed boundary temperature, and with $^{26\!}$Al as heat source (see section \ref{GTBSAnalytSol}): \citet{Miy81}, \citet{Ben96}, \citet{Tri03}, \citet{Kle08}. The basic characteristics of these models are listed in Table~\ref{GTBSTabBodyParm}. The models depend on the formatione time, $t_\mathrm{form}$, and on the radius $R$ of the parent body, and on the surface temperature $T_\mathrm{srf}$ as free parameters. The size and formation time of the body is estimated by comparing such models to meteoritic properties. For the surface temperature a value is assumed (different by different authors) that is considered as reasonable for bodies at some location in the asteroid belt.

\citet{Miy81} constructed models for the H and L chondrite parent bodies and estimated size and  formation time by comparing with maximum metamorphic temperatures of the different petrologic types and their relative abundances in observed falls and the difference in time of 100\,Ma between cooling through the closure temperature of the Rb-Sr system. The frequency of fall of the different petrologic types is now considered as unsuited for fixing parent body properties because of the highly random nature of their delivery to earth \citep[see][]{Ben96}.  \citet{Ben96} calculated models for the H and L chondrite parent bodies. The properties of the parent bodies and their formation time are estimated by comparing with maximum metamorphic temperatures of the different petrologic types and the difference in time of 60\,Ma between cooling through the closure temperature of the Pb-Pb system. \citet{Tri03} calculated models for the H chondrite parent body. Model parameters were estimated by comparison with a large set of observed cooling ages for four thermochronmeters: Ar-Ar, Pb-Pb, and Pu fission tracks retention from nine meteorites, shown in Table~\ref{GTBSTabMetDat}. The model parameters were fitted such that the resulting cooling curves for all meteorites pass as close-by as possible to the closing-temperature-closing-age data points. \citet{Kle08} calculated models for the H chondrite parent body and fitted to meteoritic data using the data set of \citet{Tri03} augmented by new data on the Hf-W thermochronometry of four meteorites (see Table~\ref{GTBSTabMetDat}). 

The weak point of the analytic models following the method of \citet{Miy81} is their inability to account for the non-constancy of the coefficients in the heat conduction equation and to add further important processes. To allow for more realistic input the basic equations have to be solved numerically, either based on the methods of finite elements \citep[used by][]{Har10} or of finite differences \citep[used by][]{Akr98,Hen12a,Hen12b,Mon13}. (There are many more studies on thermal evolution of asteroids, but without attempting to reconstruct the parent bodies of specific meteoritic classes, see the review of \citealt{Gho06}.) 

\citet{Akr98} calculated models for the H chondrite parent body that considered temperature and porosity dependent heat conductivity \citep[taken from][]{Yom83} and specific heat \citep[taken from][]{Rob79}, and in particular the insulating effect of a highly porous outer layer is considered by assuming an outer layer consisting of what they call a megaregolith layer with $30$\% porosity and an outer regolith layer with $\approx50$\% porosity.  Model parameters for the parent body were estimated by comparison with observed Pb-Pb closure ages, with metallographic cooling rates, and with maximum metamorphic temperatures of the different petrologic types.  The results show the importance of considering the insulating effect of an outer regolith layer that results in a flat inner temperature distribution and a steep descent to the surface temperature within a rather thin outer layer (see Fig.~\ref{GTBSFigCompModels}bc). This is in contrast to models assuming constant properties of the material across the body. The porous mantle  results in rather shallow burial depth's of H3 to H5 chondrites while most of the interior is of type H6 (see Fig.~\ref{GTBSFig6Hebe}).

\citet{Har10} calculated models for the H chondrite parent body based on essentially the same approximations as in \citep{Akr98}. The model parameters, including thicknesses of regolith and megaregolith layers, are determined such as to obtain an optimum fit to cooling age data of eight meteorites (see Table~\ref{GTBSTabMetDat}, without Mt. Browne), metallographic cooling rates for twenty meteorites, and cooling rates determined from Pu fission-tracks for seven meteorites. Also compared are the maximum temperature at the burial depths with (assumed) maximum metamorphic temperatures of the petrologic type.  

\citet{Hen12b,Hen12a} calculated models for the H chondrite parent body including a  modelling of the evolution of porosity by solving the equation for compaction by cold pressing \citep{Gue09} and the equations for sintering of granular material by hot pressing at elevated temperatures similar to \citep{Yom84}. In this model the body develops a compacted core, but retains a surface layer of significant porosity (see Fig.~\ref{GTBSFigSinterPor}), analogue to \citet{Akr98} and \citet{Har10}, but now determined self-consistently from the sintering process. The model parameters were determined by a least square method that minimizes the distances between a cooling curve and the corresponding data points and an automated search strategy for the model parameters based on a variant of the genetic algorithm \citep{Cha95}. 

In \citet{Hen13} the model was somewhat improved. It was found that the best fit between models and meteoritic data is obtained with a rather low $^{60}$Fe abundance of the order of $10^{-8}$ for the $^{60}$Fe/$^{56}$Fe abundance ratio, that is close to the latest value of $5.9\pm1.3\times10^{-9}$ for the  $^{60}$Fe/$^{56}$Fe ratio found by \citet{Tan12}. In some models (see Table \ref{GTBSTabBodyParm}) $^{60}$Fe is considered as a heat source but at the low abundance now assumed, $^{60}$Fe does not substantially contribute to the heating of planetesimals and can be neglected. ´The model was also extended to include a finite growth time of the parent body of the H chondrites by considering mass-influx over some finite period and introducing a moving outer boundary. It turned out that only short accretion times of up to 0.3\,Ma are compatible with the empirical data on the cooling history of H chondrites. The instantaneous accretion hypothesis widely used in model calculations is supported by this finding.
 
\begin{figure*}[t]

\centerline{
\includegraphics[width=\hsize]{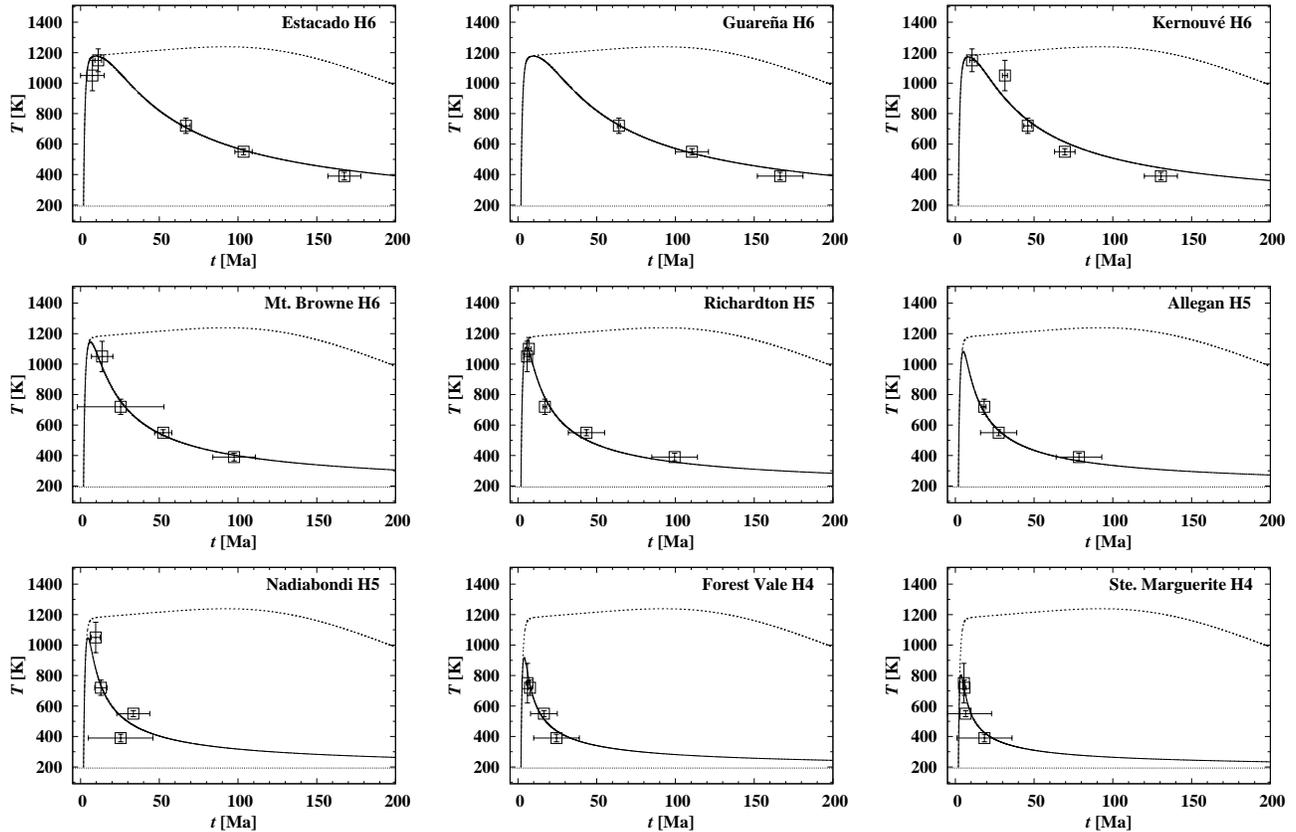}
}

\caption{\small Thermal evolution history of individual meteorites in an optimized evolution model for the parent body of H chondrites. The values of the free parameters found for the optimised model of \citet[][ Table 6]{Hen12a} are given in Table~\ref{GTBSTabBodyParm}, model 'Hen12'. The solid lines shows the temperature evolution at the burial depth's of the meteorites, the dashed lines the temperature evolution at the centre of the asteroid, and the dotted lines corresponds to the surface temperature. The square boxes with error bars indicate the individual data points and the estimated uncertainty of closing temperature and age determinations.}
\label{GTBSKEvolMetOpt}
\end{figure*}

\citet{Mon13} calculated models for the H chondrite parent body based on similar approximations as in \citep{Akr98}. A power-law temperature dependence of heat conductivity was assumed. A growth of the body was allowed for by the method of a moving grid as in \citet{Mer02}. The same set of thermochronological data as in \citet{Hen12b} was used for comparison. A big number of models were calculated for widely varying parameter values to determine the region in parameter space inside of which a close fit of the meteoritic cooling histories is possible.  The consequences of an extended accretion period were studied and it turned out also in this model calculation that good fits to the meteoritic record can only be obtained for growth periods less than about 0.2\,Ma.

Figure \ref{GTBSFigCompModels} shows for comparison the thermal structure of some ``standard'' body of a fixed mass that corresponds to a fully consolidated body of 100\,km radius with the composition of the H chondrite parent body. The heating rate is also assumed to be the same in all models in order that the total amount of heat released is transferred to the same mass in all cases considered (for details see figure caption). The thermal structure is shown for models based on the different kinds of approximations used so far for modelling the parent body of the H chondrites. One recognizes two important points: (i) models with a constant heat capacity as assumed result in much too high temperatures, and (ii) models without insulating outer layers result in much too low temperatures in the shallow outer layers. A correct treatment of the heat capacity and inclusion of an outer regolith layer are indispensable for realistic models. 

However, despite of the different assumptions made in the model calculations, from the rather simplistic model of a homogeneous body with constant material properties, that can be treated analytically \citep{Miy81}, to models with rather complex implemented physics like that of \citet{Har10} and \citep{Hen12b}, the derived properties of the parent body are rather similar (see Table~\ref{GTBSTabBodyParm}). These do not seem to depend very critical on the particular assumptions. The more strongly deviating results for the burial depths of the meteorites found in the different models cannot be used to discriminate between models, because no independent information on this is presently available. This would require that methods are developed to determine the maximum pressure experienced by the material at the rather low pressures in  meteoritic parent bodies. 

There are also a number of other model calculations \citep[e.g.,][]{Ben96,Hev06,Sah07,War11} that studied the thermal evolution of asteroids and gave important insight into their properties and the processes operating in them, but did not attempt to obtain definite results for the properties of the parent body of H chondrites from some kind of ``best fit'' to meteoritic data. Only the range of possible values is studied in these papers. This does not enable a direct comparison, but the range of possible values for radii and formation times given are in accord with the results shown in Table~\ref{GTBSTabBodyParm}.

\bigskip
\noindent
\textbf{\refstepcounter{subsection}\thesubsection\  Optimized fit to meteoritic data}
\bigskip

A systematic method to fit thermochronological data with results of model calculations for the thermal evolution of planetesimals was applied by \citep{Hen12b,Hen13}. This method allows to use a large number of input data of meteorites and to deal with a large number of parameters. For this purpose the code for modelling the thermal evolution of planetesimals is coupled with a code for minimum search in order to obtain optimized fits of thermal evolution models to a set of empirical data for the cooling history of meteorites. The quality function to be minimized corresponds to the $\chi^2$ method; details are described in \citet{Hen12b}. As method for the minimum search the so called genetic algorithm in the variant of the method described by \citet{Cha95} was chosen. This mimics the basic principles that are thought to rule biological evolution of species and is thought to find optimal values even under extremely unfavourable conditions. This automated search for a minimum allows to perform much more complex comparisons with meteoritic data (bigger numbers of meteorites, bigger number of parameters, inclusion of meteorites with incomplete or less-quality data) than is possible if the optimisation is done 'by hand' as in previous works. 
 
The method was applied to the case of H chondrites because there are nine chondrites (see Table~\ref{GTBSTabMetDat}) that satisfy the requirements that their characteristics suggest that they could have a common origin from one parent body, that at least three data points are available for each of them, and that they are of shock stage S0, such that no reset of the thermochronologic clocks is expected. For these meteorites totally 37 data points are available while the parent body model depends on four parameters (radius $R$, formation time $t_{\rm form}$, surface temperature $T_{\rm srf}$, initial porosity $\Phi_{\rm srf}$). Because the measured bulk heat conductivities $K_{\rm b}$ of chondrites shows large scatter \citep{Yom83} also this quantity was considered as free parameter that is included in the optimisation process. Additionally one has the nine unknown burial depth's of the meteorites as parameters. Totally, the problem requires to fit the 37 data points with a model depending on 14 parameters. It turns out that a very good fit is possible resulting in a value of $\chi^2<7$. The quality of the fit can be considered as excellent, because an acceptable fit only requires a value of $\chi^2<23$ (= number of data (37) less number of parameters (14)).

Figure \ref{GTBSKEvolMetOpt} shows the evolution of the temperature at the burial depth for the optimized model for the meteorites and the corresponding data points for closure time and closure temperature. Figure \ref{GTBSFig6Hebe} shows a schematic graphical representation of the burial depths of all the meteorites used for the optimisation process. The results demonstrate that it is possible to find a solution for the properties of the putative common parent body of these meteorites and for their burial depths that satisfactory reproduces all experimental findings. 

This finding lends some credit to the onion shell model to explain the petrologic types of the meteorites. It would be desirable to extend such calculations to other meteoritic classes, but this requires that more thermochronological data are acquired.

\bigskip
\begin{minipage}{\hsize}
\centerline{\textbf{\refstepcounter{section}\thesection. \label{GTBSVesta}
VESTA, THE PARENT BODY OF THE}}
\centerline{\textbf
HED METEORITES}
\end{minipage}
\bigskip

The howardite-eucrite-diogenite (HED) meteorites are a subgroup of the achondrites and originate from a differentiated parent body that experienced extensive igneous processing.  The spectral similarity between these meteorites and the asteroid 4 Vesta indicated their genetic link  \citep[e.g., ][]{McC70} and recently the Dawn mission that orbited Vesta between July 2011 and September 2012, has supported the conjecture of Vesta as the parent body. HED's consist mainly of noncumulative eucrites (pigeonite-plagioclase basalts) and diogenites that are composed of orthopyroxene and olivine of different mixtures. Howardites are impact breccias, composed predominantly of eucrite and diogenite clasts. The igneous lithology is completed with more rare rock types that include cumulative eucrites. An eucritic upper crust and orthopyroxene-rich lower crustal layer have been identified by Dawn in the Rheasilvia basin  \citep{DeS12,Pre12}. The huge basin allows the identification of material that was formed down to a depth of 30 -- 45\,km and has been excavated or exposed by an impact \citep{Jut11,Iva13}. Olivine-rich and thus mantle material has not been found in a significant amount, i.e., not above the detection limit of $<25$\%, suggesting a crustal thickness of at least 30 -- 45\,km  \citep{McS13}. This finding may also explain the lack of olivine-rich samples in the HED collection.

The lithology of the HED's described above has been used to constrain the interior structure of the parent body. Vesta is believed to have a layered structure consisting of at least a metallic core, a rocky olivine-rich mantle and a crust consisting mainly of an upper basaltic (eucritic) and a lower orthopyroxene-rich layer  \citep[e.g., ][]{Ruz97,Rig97,Man13}. The details in the interior structure, i.e., core size, mantle and crust thickness, however, vary between the models and depend among others on the assumption of the unknown bulk composition of Vesta.  However, the recent gravity measurements from the Dawn mission have provided a further constraint on the interior structure, i.e., the core radius with a size of 107 -- 113\,km assuming core densities between 7\,100 and 7\,800 kg m$^{-3}$ \citep{Rus12}. Some HEDs, furthermore, show a remnant magnetisation suggesting a formerly active dynamo in a liquid metallic core \citep{Fu12}.

To infer the differentiation events of Vesta, various radioisotopic dating systems have been used. From these data it has been concluded that the core-mantle differentiation likely precedes the mantle-crust differentiation. Eucrites exhibit siderophile depletions that indicate the formation of an iron-rich core within $\approx1$ to 4\,Ma of the beginning of the solar system and prior to their crystallization \citep{Pal81,Rig97,Kle09}. The Hf-W ages suggest a crystallization time for eucrite zircons with up to $\approx7$\,Ma after CAI formation whereas for eucrite metals it is $\approx20$\,Ma \citep{Sri07}. The latter, however, may reflect impact-triggered thermal metamorphism in the crust of Vesta \citep{Sri07}. In contrast to these findings, the $^{26}$Mg composition of the most primitive diogenites requires the onset of crystallization already within  0.6 $\pm$ 0.4\,Ma after CAI's and the variation in $^{26}$Mg in diogenites and eucrites indicates complete solidification, i.e., rapid cooling of the body, even within the first 2 -- 3\,Ma after CAI's \citep{Schi11}.  

Two possible differentiation scenarios have been associated with the formation of the basaltic achondrites (i.e., eucrites and diogenites). The first scenario suggests that eucrites and diogenites originated from the partial melting of the silicates  \citep[e.g., ][]{Sto75,Sto77,Jon84}  with the extraction of basaltic (euritic) magma leaving behind a harzburgite, orthopyroxenite or dunite residual depending on the degree of partial melting. The other scenario favored by geochemical arguments suggests achondrites being cumulates formed by magma fractionation. In that scenario the diogenites could have crystallized either in a magma ocean  \citep[e.g., ][]{Ike85,Rig97,Ruz97,Tak97,War97,Dra01,Gre05,Schi11} or in multiple, smaller magma chambers  \citep[e.g., ][]{She97,Bar08,Bec10,Man13}. Eucrites are then products from the magmas that had earlier crystallized diogenites.

Several numerical and experimental studies  \citep[e.g., ][]{Gho98,Rig97,Dra01,Gup10,Elk11,For13,Neu13} have been performed in order to model Vesta's thermal and geological evolution. As a general conclusion from the thermal models one can state that Vesta should have finished its accretion by $\approx1.5$ -- 3\,Ma after the CAIs to efficiently differentiate due to the decrease of the radiogenic heating by $^{26\!}$Al and $^{60}$Fe with time. The timing and duration of crust and core formation then vary between the models in particular as in most cases differentiation processes have not been modelled self-consistently; rather some specific scenario has been assumed and its consequences for the thermal evolution have been studied.  For instance,  \citet{Rig97} considered core formation and crystallisation of a cooling magma ocean using a numerical physicochemical model. The core is assumed to have separated during a global magma ocean episode at a melt fraction of 65 to 77\%.  Then after cooling to a crystal fraction of 80\%, residual melt percolated from the former extensive magma ocean to form eucrites at the surface and diogenites in shallow layers by further crystallisation. In their model complete crystallisation occurred within 20\,Ma after the formation of Vesta.   \citet{Gup10} performed a thermal model and considered convective cooling in a global magma ocean. They investigated two evolution paths, i.e., the formation of basaltic achondrites via partial melting of silicates or as residual melts after crystallisation of a convecting magma ocean. For the partial melting model, they concluded that, depending on the formation time, melt extraction is possible between 0.15 and 6\,Ma after the CAIs and that differentiation proceeds rapidly within O(10$^4$) a. For the scenarios where accretion was completed within 2\,Ma after the CAIs, a magma ocean formed, which does not crystallise completely for at least 6 -- 10\,Ma.  \citet{Gho98} investigated the differentiation of Vesta by assuming instantaneous core formation in the temperature interval of 1\,213 -- 1\,223 K and that HED meteorites are the product of 25\% partial melting. To obtain such a scenario, they concluded that Vesta must have accreted at 2.85\,Ma, differentiated at 4.58\,Ma and formed a basaltic crust at 6.58\,Ma relative to the formation of the CAIs. Furthermore, they suggested that the mantle remained partially for 100\,Ma after its formation and that some near-surface layers may have remained undifferentiated. A recent study by  \citet{For13} considers core formation by porous flow of iron melt. Core differentiation takes place in all scenarios in which Vesta completes its accretion in less than 1.4\,Ma after CAI's and reaches 100\% of silicate melting throughout the whole asteroid if Vesta formed before 1\,Ma. The model, however, neglects volcanic heat transport, convection in the magma ocean and redistribution of radioactive heat sources. Considering those as well, results to a different evolution path: Efficient cooling by melt migration and redistribution of $^{26\!}$Al toward the surface inhibits the formation of a global magma ocean, instead a shallow magma ocean forms close to the surface  \citep{Neu13}.

Thus, whether the basaltic crust did form after the silicate-iron separation due to the crystallization of a global magma ocean, or prior to the silicate-iron separation due to the extrusion of the first partial melts (partial melting origin or multiple, smaller magma chambers), remains an open issue. Which scenario is more likely depends on the competition between heating by short-lived isotopes and the depletion of these heat sources due to their partitioning into the melt and the migration velocity of this melt (see discussion above in section~\ref{GTBSSectDiffer} about global magma ocean). Only if the extrusion of melt containing $^{26\!}$Al is comparatively slow, an asteroid-scale magma ocean may have formed otherwise HED's formed by the extrusion of early partial melts that are either collected in smaller magma chambers or produced a shallow rapidly (within O(10$^4$-10$^6$\,Ma) freezing magma ocean close to the surface \citep{Neu13}. The latter models also suggest that mantle-crust differentiation precedes core formation, which seems to contradict the chronological records of HED's seem indicating a reversed sequence of differentiation \citep{Kle09}.

On the other hand, a global magma ocean is difficult to reconcile with the chronological records of the basalts that suggest their early origin as a global magma oceans is suggested to produce residual basaltic melts much later than during the first 10 million years  \citep{Gup10,Schi11}. The shallow magma ocean model  \citep{Neu13} is consistent with an early formation time of $t_0<1$\,Ma of diogenites and the rapid cooling of the body, even within the first 2 -- 3\,Ma after CAI's \citep{Schi11}.  It further implies that the first non-cumulate eucrites form before diogenites, consistent with internal  $^{26\!}$Al -- $^{26}$Mg isochrones recently obtained by \citet{Hub13} that infer younger ages for the diogenites compared to those of the eucrites, and younger ages for the cumulative eucrites compared to the non-cumulative ones.

\bigskip
\textbf{Acknowledgments.} 
This work was supported by `Schwerpunktprogramm 1385' of the `Deutsche 
Forschungs\-gemeinschaft (DFG)'. We aknowledge that W.~H. Schwarz and S. Henke made available some of the figures.



\end{document}